\documentclass[12pt]{article}
\usepackage{comment}
\usepackage{enumitem} 
\usepackage[utopia]{mathdesign}
\usepackage[colorlinks,citecolor=blue]{hyperref}
\usepackage{authblk}
\usepackage[normalem]{ulem}
\usepackage{tikz}
\usepackage{nameref}
\usepackage{changepage}
\usepackage{amsmath, graphicx}
\usepackage{float}
\usepackage{caption}
\usepackage{subcaption}
\usepackage[export]{adjustbox} % Optional: for 'valign=t' if needed and easy width control

\usepackage[onehalfspacing]{setspace}  
\usepackage{sectsty}
\sectionfont{\centering \Large \scshape }
\subsectionfont{\centering \large  \scshape}
\usepackage{geometry}
\usepackage{marginnote}
\usepackage{xcolor}
\usepackage{amsthm}

\usepackage[title]{appendix}


\newcounter{parentnumber}

\theoremstyle{plain}          % Sets italic body text, bold header
\newtheorem{corollary}{Corollary}          % Creates: Corollary 1, Corollary 2, ...
\theoremstyle{definition}     % Sets normal text body, bold header  
\theoremstyle{plain}          % Back to italic body text, bold header
\newtheorem{lemma}{Lemma}                  % Creates: Lemma 1, Lemma 2, ...
\newtheorem{observation}{Observation} 
\newtheorem{proposition}{Proposition}      % Creates: Proposition 1, Proposition 2, ...
\newtheoremstyle{mystyle1}%
  {\topsep}%                  % Space above theorem
  {\topsep}%                  % Space below theorem  
  {\itshape}%                 % Body font: italic
  {}%                         % Indent amount (empty = no indent)
  {\bfseries \scshape}%       % Header font: bold + small caps
  {.}%                        % Punctuation after header
  {5pt plus 1pt minus 1pt}%   % Space after header (flexible)
  {}%                         % Custom header specification (empty = use default)

\theoremstyle{mystyle1}
\newtheoremstyle{mystyle2}%
  {\topsep}%                  % Space above theorem
  {\topsep}%                  % Space below theorem
  {{\normalfont}}%            % Body font: normal (not italic)
  {}%                         % Indent amount (empty = no indent)
  {\bfseries \scshape}%       % Header font: bold + small caps  
  {.}%                        % Punctuation after header
  {5pt plus 1pt minus 1pt}%   % Space after header (flexible)
  {}%                         % Custom header specification (empty = use default)

\theoremstyle{mystyle2}
\usepackage[american]{babel}
\usepackage{csquotes}
\usepackage[style=apa, uniquelist=false, backend=biber, doi=false,isbn=false,url=false]{biblatex}
\DeclareLanguageMapping{american}{american-apa}

\title{\vspace{-4em}\Large Public Good Provision with a Governor\thanks{\textbf{Conflict of Interest}: None} 
\thanks{We are grateful to the Editor, Associate Editor and two anonymous referees for their insightful and constructive comments. We also benefited from the feedback of Jorge Bruno, Trivikram Dokka, Renaud Foucart, Konstantinos Georgalos, Tin Leelavimolsilp, and Orestis Troumpounis, as well as from seminar and conference participants in Bath, Trento, and Lancaster.}}

\author{ \normalsize Chowdhury Mohammad Sakib Anwar\thanks{\textit{Author for correspondences.} Faculty of Business and Digital Technology, University of Winchester. Email: sakib.anwar@winchester.ac.uk}
\and \normalsize  Alexander Matros\thanks{Darla Moore School of Business, University of South Carolina. Email: alexander.matros@gmail.com}
\and \normalsize  Sonali SenGupta\thanks{Department of Economics, Queen's Business School, Queen's University Belfast. Email: s.sengupta@qub.ac.uk }}

\date{{\normalsize  \today} \\ 
\normalsize}

\begin{document}

\maketitle
\begin{abstract}

We study a public good game with $N$ citizens and a Governor who allocates resources from a common fund. Citizens may voluntarily contribute or be compelled to do so if audited, in which case shirkers face a penalty. The Governor decides how much of the fund to devote to public good provision, with the remainder embezzled. Crucially, the Governor’s utility combines material payoffs from embezzlement with belief-dependent reputational concerns. We characterize the symmetric subgame perfect equilibria (SSPE) of the game. The model always admits at least one pure-strategy equilibrium, ranging from universal free-riding with complete embezzlement to full contribution with efficient provision. 
Symmetric mixed-strategy equilibria arise only within a limited region of the parameter space and may not be unique.
Our analysis highlights the roles of penalties, audits, and reputational incentives in sustaining contribution and provision, thereby linking public good provision with the broader literature on corruption, embezzlement, and psychological game theory.

\medskip
\begin{flushleft}\textbf{Keywords} : {Public goods; Psychological game theory; Embezzlement; Governor.}\end{flushleft}\par
\begin{flushleft}\textbf{JEL Codes}: D73, H40 \ \end{flushleft}\par
\end{abstract}
\newpage

\section{Introduction}

Provision of public goods often involves delegation to intermediaries: be they bureaucrats, governors, or aid administrators, who collect resources from citizens but wield considerable discretion over their final allocation. This creates the potential for embezzlement, especially when the intermediary's actions are only partially observable \parencite{ferraz2008exposing,Reinikka2004}. Consider, for instance, the recent high-profile case in Mauritius: the country's former central bank governor and finance minister were arrested over alleged embezzlement of public funds meant to support COVID-19 affected businesses \parencite{Reuters2025Mauritius}. Yet, such episodes also highlight how public expectations and reputational concerns may constrain misappropriation by intermediaries, creating a tension between material incentives and psychological costs of guilt. This raises a crucial, but under-explored, theoretical question: how do limited citizens' auditing and an intermediary's concern for public expectations jointly shape the incentives for citizens to contribute, and for the governor to redistribute public funds to provide public goods? 

%This paper introduces a tractable three-stage model incorporating citizen contribution decisions, random auditing, and a provision decision by a potentially corrupt governor.  A finite number of identical citizens decide whether to contribute to a common fund, knowing that a fixed number of them will be audited and compelled to contribute if found shirking, with an additional penalty.  The governor observes only the total fund and acts in self-interest to choose how much to allocate to the public good, keeping the remainder as her private rent. While she cannot be punished for embezzlement, her utility is belief-dependent and she suffers a reputational loss if provision falls short of the public’s expectations. These expectations are formed rationally from the citizens’ contribution strategies. 

This paper introduces a tractable three-stage model incorporating citizen contribution decisions, random auditing, and a provision decision by a potentially corrupt governor. A finite number of identical citizens decide whether to contribute to a common fund, knowing that a fixed number of them will be audited and compelled to contribute if found shirking, with an additional penalty. The governor observes the total fund and, acting in self-interest, chooses how much to allocate to the public good, keeping the remainder as private rent. Citizens observe the public good provided by the governor but not the size of the common fund. While the governor cannot be punished for embezzlement, her utility is belief-dependent: she suffers a reputational loss if provision falls short of what she believes the public expects, where expectations are formed rationally from citizens’ contribution strategies.

%This paper develops a simple but rich model capturing this interplay between citizens’ contribution decisions, probabilistic monitoring, and a distributor’s allocation choice under reputational concerns. A finite number of identical citizens decide whether to contribute to a common fund, knowing that a fixed number of them will be audited, and compelled to contribute if found shirking, with an additional penalty. The Governor observes only the total fund and \textcolor{red}{acts in self-interest to} choose how much to allocate to the public good, keeping the remainder as her private rent. While she cannot be punished for embezzlement, she suffers a reputational loss if provision falls short of the public’s expectations. These expectations are formed rationally from the citizens’ contribution strategies.

Our analysis yields three key contributions. First, we provide the conditions under which cooperation emerges (full contribution equilibria), collapses (free-riding equilibria), or is sustained in mixed-strategy equilibria where citizens randomize between contributing and shirking. %First, we characterize the full set of symmetric subgame perfect equilibria, including free-riding equilibria \textcolor{red}{(where no one contributes)}, full-contribution equilibria (where everyone contributes), and mixed-strategy equilibria where citizens randomise between contributing and shirking. 
Second, we derive explicit parameter thresholds: in terms of the penalty, audit intensity, and the governor’s reputation sensitivity, that separate these equilibria, and illustrate them with complete parameter maps. Governor's reputational sensitivity is linked with citizens' expectations, which are always correct and self-enforced in the symmetric equilibria. Third, we show how reputational concerns interact with enforcement: high penalties can sustain contributions even without reputational sensitivity, but when penalties are low, reputation can substitute as a disciplining device. We illustrate our model and general results with a simple example and extensive form game-tree, involving two citizens and a governor, with citizens facing equal probability of being audited. In this simple framework, we make the following observations: \textit{(i)} shirking is optimal when expected penalty is low, and governor provides low or zero public funds, depending on her reputation sensitivity; \textit{(ii)} each citizen is pivotal to ensure an efficient public good provision; \textit{(iii)} there exists a unique or continuum of mixed-strategy equilibria, depending on the return from public good and penalty if caught shirking.

This work connects to the public goods literature on enforcement and punishment (e.g., monitoring in voluntary contribution games), the study of corruption and embezzlement in resource allocation, and behavioural models of guilt aversion, image concerns, and reciprocity, stemming from the literature on psychological game theory (PGT). By integrating these elements into a single framework, we provide both analytical characterizations and practical insights for policy design, in particular, on the joint calibration of penalties and monitoring intensity to maintain high provision levels, despite the risk of embezzlement.

%Our work is situated at the intersection of three extensive literatures: public goods, the economics of corruption, and Psychological Game Theory (PGT). 

We explicitly frame our analysis within PGT \parencite{battigalli2022belief}, as the governor's utility, which incorporates a direct utility loss if she fails to meet citizens' expectations, is a form of guilt aversion \parencite{CharnessDufenberg2006, BattigalliDufenberg2007}, and reputational concerns \parencite{BenabouTirole2006, EllingsenJohannesson2008}. This aligns our model with theoretical and experimental work showing that agents are motivated to live up to the expectations of others, thereby avoiding the psychological cost of guilt \parencite{dufwenberg2004theory, ellingsen2010testing}. Our primary contribution is to embed this PGT mechanism within a public goods setting to analyse its interaction with traditional enforcement tools.

 We build on the classic public goods literature that studies mechanisms to overcome free-riding, where much of the work shows the effectiveness of centralised sanctioning \parencite{Baldassarri2012} and (more) successful informal punishment mechanisms like peer punishment \parencite{fehr2000Gachter, fehr2002altruistic}.\footnote{Some relevant articles on centralised sanctioning include  \textcite{Chaudhuri2011, ledyard95, andreoni1996governmen, Markussen2014, Markussen2016}. In addition, recent literature on sequential public good provision games with groups, positional uncertainty and observational learning,  emphasize the pivotal role each agent plays within a group to encourage contributions, and how a fully cooperative equilibrium can be sustained without the need for punishment. See \textcite{anwar2025efficient}, and a special case with individual agents \parencite{gallice2019co}, and a related experiment \parencite{AnwarGeorgalos2024}.} Our model departs from the standard framework by analysing this mechanism not in isolation, but in the presence of a self-interested governor who controls the final allocation. This approach connects our paper to the literature on reciprocal obligations: citizens' contribute and the governor provides public goods, wherein there is co-evolution of civic culture and the intrinsic reciprocity between citizens and the governor (see \textcite{besley2020state,besley2023norms}).
%
%Our work is therefore related to models of the co-evolution of civic culture and state capacity, such as \textcite{besley2020state}. \textcite{besley2020state} models the long-run evolutionary dynamics of a society with heterogeneous citizens, where as our paper provides a single-period sequential model where the public goods provision depends on interaction between homogeneous citizens and a distributor.

Our paper contributes to the broader literature on political corruption and embezzlement \parencite{ades1999rents, brollo2013political}. Our focus on a non-electoral disciplining device distinguishes our work from the extensive literature on political accountability, where the primary mechanism for controlling corruption is the threat of losing an election \parencite{ferraz2011electoral,weitz2017can,welch1997effects}. For instance, unlike \textcite{van2013candidates}, who model re-election incentives between candidates with known policy preferences, we model a form of social accountability where citizens' expectations act as a direct constraint on governor's behaviour. We contribute a complementary perspective by modelling this form of social accountability, where citizens' expectations act as a direct, non-electoral constraint on the governor's behaviour, which can be interpreted as an aversion to triggering social sanctions like whistleblowing, tip-offs, protests, filing lawsuits, formal complaints, as documented in the literature on "grievance redress" \parencite{worldbank_2013, fox2015social}.\footnote{\textcite{TianZhao2024} provide experimental evidence on how the attitude of citizens affected by the negative externalities of corruption significantly influences levels of corruption. A related experimental study with a stylized sequential embezzlement game, combining ultimatum game and common resource game, studies how organisational structure influences embezzlement \parencite{MakowskyWang2018}. This is also distinct from the formal persuasion game with limited verification in \textcite{lambert2015social}.}

We also connect our work to the literature on tax evasion \parencite{allingham1972income, alm2019motivates}. 
Few relevant papers analyse the interplay between corruption, tax evasion and public goods (see \textcite{Litina2016} and \textcite{lambert2015social}). Our primary contribution to this nascent literature is our tractable framework that explicitly models the strategic interaction between citizen contribution, and governor's reputational concerns. Unlike these related models, our framework allows for a complete characterisation of different equilibria including free-riding, full contribution, and mixed strategies, and the precise conditions under which each will emerge, offering clear, policy-relevant insights into the trade-offs between formal punishment and reputational incentives.\footnote{Our work is also related to a fairly recent literature on morality of agents in social dilemma situations: see \textcite{vanLeeuwenAlger2024, subrado2022moralagents, AlmTorgler2011, EichnerPethig2021}, and the references within.}

Finally, our theoretical results provide a bridge to the experimental literature on embezzlement and guilt aversion. For example, \textcite{attanasi2019embezzlement} provide experimental evidence that guilt aversion can reduce embezzlement by intermediaries. Our model complements these findings by generating a rich set of sharp, testable predictions about how the magnitude of this effect should vary with the intensity of auditing and the severity of formal penalties, thus offering a clear path for future experimental investigation.

The remainder of the paper is organized as follows. Section \ref{sec:model} presents the model. Section \ref{sec:illustrative} presents an illustrative example. Section \ref{sec:analysis} characterize symmetric subgame perfect equilibria (SSPE), including the efficient equilibrium in which all citizens contribute voluntarily and the Governor allocates the entire common fund to public good provision. Section \ref{sec:mixed} analyzes mixed-strategy SSPE.  Section \ref{sec:conclusion} concludes the paper.

\section{The Model} \label{sec:model}

We analyse a three-stage game of imperfect information between a population of $N \ge 2$ identical citizens, indexed by $i \in \mathcal{I} = \{1, \dots, N\}$, and a Governor, $G$. 
The sequence of actions and the information structure of the game are as follows:

\paragraph{Stage 1: Citizens' Contribution Decision.}
Each citizen $i \in \mathcal{I}$ simultaneously chooses an action $c_i \in \{0,1\}$, where $c_i=1$ represents contributing one monetary unit to a common fund and $c_i=0$ represents shirking. Citizens make this choice without observing the actions of others. The collection of choices $(c_1, \dots, c_N)$ constitutes the contribution profile, which is
not observed by any other player, including the Governor.

\paragraph{Stage 2: Auditing by Nature.}
After contributions are chosen, Nature (a non-strategic player) randomly selects exactly $k \in \{1, \dots, N\}$ 
%\textcolor{red}{(should it be $k \in \{1, \dots, N-1\}$)} 
distinct citizens for an audit, sampling without replacement. The probability that any given citizen $i$ is selected for an audit is therefore $k/N$. Let $\omega_i \in \{0,1\}$ be an indicator variable where $\omega_i=1$ if citizen $i$ is audited and $\omega_i=0$ otherwise. The audit profile is $(\omega_1, \dots, \omega_N)$, with $\sum_{i=1}^N \omega_i = k$. The realisation of $\omega_i$ is private information to citizen $i$. No other player, including the Governor, observes the audit profile.

If a citizen shirks ($c_i=0$) and is audited ($\omega_i=1$), he must pay the one-unit contribution as well as an additional penalty $z \ge 0$. We assume this penalty represents a pure deadweight loss and does not contribute to the common fund.

%If a citizen who shirked ($c_i=0$) is audited ($\omega_i=1$), they are compelled to pay the one-unit contribution and an additional penalty $z \ge 0$. We assume this penalty is a deadweight loss and does not accrue to the common fund.

\paragraph{Stage 3: Governor’s Allocation Decision.}

The total common fund $X$ consists of all voluntary contributions and any contributions compelled through audit:  
\begin{equation} \label{eq:X}
    X = \sum_{i=1}^N \max \{ c_i, \omega_i\}.
\end{equation}

The Governor observes the common fund $X$, but not the individual contribution profile $(c_i)_{i \in \mathcal{I}}$ nor the audit profile $(\omega_i)_{i \in \mathcal{I}}$. 
%Based solely on the common fund $X$, the Governor
She chooses a provision level for the public good, $g \in [0,X]$. The residual amount, $X-g$, is embezzled for her private benefit.  Citizens, in turn, observe only the realized provision level $g$; they do not observe the size of the common fund $X$.

Throughout the analysis, the audit intensity $k$ and the penalty $z$ are treated as exogenous institutional parameters, while the Governor’s only strategic decision is the allocation of the realized common fund.
Our solution concept is a Subgame Perfect Equilibrium (SPE).

%\textbf{\textit{A pure strategy for a citizen is a contribution decision $c_i \in \{0,1\}$. We focus on symmetric mixed-strategy profiles in which all citizens contribute with the same probability $p \in [0,1]$.  }}

%\textbf{\textit{Given such a symmetric profile, citizens form a belief about the expected size of the common fund, denoted by $\tau(p)$. This expectation plays a central role in the Governor’s decision problem and will be derived explicitly in the analysis below.}}

The utility for each player is defined as follows.

\paragraph{Citizens.} The utility of citizen $i$ is linear in his monetary payoff and is given by:
\[
u_i(c_i, \omega_i, g) = - \max\{c_i, \omega_i\} - (1-c_i)\omega_i z + ag,
\]
where $a \in (0,1)$ is a parameter that captures the marginal per capita return from the public good. The first term is the citizen's contribution (voluntary or compelled), the second is the penalty if caught shirking, and the third is the benefit from public good provision.

%\textcolor{red}{We need to specify how we are defining the utility in general and then go into our specific case (i.e. symmetric case)}. \textcolor{green}{Done! see below}

\paragraph{Governor.}

The Governor's utility consists of two components:  
\begin{enumerate}[label=(\roman*)]
    \item a \emph{material payoff}, given by the residual funds embezzled, $X-g$; and  
    \item a \emph{reputational payoff}, or equivalently a psychological cost of guilt, which can be interpreted as political capital, public trust, guilt aversion \parencite{ellingsen2010testing}, or reciprocity \parencite{dufwenberg2004theory}. 
    This component is determined by the discrepancy between the realized public good provision $g$ and the level that the Governor believes citizens expect to observe.
\end{enumerate}

Let $\tau(p_1,\ldots,p_N)$ 
denote the level of the common fund that the Governor believes citizens expect to observe, where
$p_i \in [0,1]$
is the probability with which citizen 
$i$ contributes.

%\textbf{\textit{A pure strategy for a citizen is a contribution decision $c_i \in \{0,1\}$. We focus on symmetric mixed-strategy profiles in which all citizens contribute with the same probability $p \in [0,1]$.  }}

%Let $\tau(p)$ denote 
%\color{red}{the level that the Governor believes citizens expect to observe} 
%in the common fund. 

%The Governor’s utility is then  
%\[
%u_G(g, \tau(p_1, ..., p_N), X) = (X - g) - E_{PG}(g, \tau(p_1, ..., p_N)).
%\]

%\textbf{\textit{We focus on symmetric mixed-strategy profiles in which all citizens contribute with the same probability $p \in [0,1]$.  }}
%Under symmetric voluntary contributions with probability $p$, the Governor believes citizens expect to observe the following common fund:

%\[
%\tau(p, ..., p) = \tau(p) = \mathbb{E}[X].
%\]

\iffalse

We model the reputational cost via a step-loss function:  
\[
%E_{PG}\!\left(g, \tau(p_1, ..., p_N)\right)
%= 
\sigma \cdot \mathbf{1}\{g < \tau(p_1, ..., p_N)\}
= \begin{cases}
    \sigma, & g < \tau(p_1, ..., p_N), \\[6pt]
    0, & g \geq \tau(p_1, ..., p_N),
  \end{cases}
\]
where $\sigma > 0$ measures the Governor’s sensitivity to falling short of expectations.  

\fi

The Governor’s utility is then

\begin{equation}\label{eq:gov_util_a}
   u_G(g, \tau(p_1, ..., p_N); X, \sigma) = (X - g) - \sigma \cdot \mathbf{1}\{g < \tau(p_1, ..., p_N)\},
\end{equation}

where we model the reputational cost via a step-loss function:  
\[
%E_{PG}\!\left(g, \tau(p_1, ..., p_N)\right)
%= 
\sigma \cdot \mathbf{1}\{g < \tau(p_1, ..., p_N)\}
= \begin{cases}
    \sigma, & g < \tau(p_1, ..., p_N), \\[6pt]
    0, & g \geq \tau(p_1, ..., p_N),
  \end{cases}
\]
where $\sigma > 0$ measures the Governor’s sensitivity to falling short of expectations. We define the Governor to be \textit{reputation-sensitive} if $\sigma \geq  \tau(p_1,...,p_N)$, and \textit{reputation-insensitive}, otherwise.

The Governor observes only the total common fund and chooses how much to allocate to the public good, balancing material incentives from embezzlement against a belief-dependent reputational cost if provision falls short of citizens’ expectations.

%Our solution concept is a Subgame Perfect Equilibrium (SPE). 

\paragraph{Beliefs and expectations.}

Belief-dependent preferences in the model are attributed exclusively to the Governor. 
Citizens’ utility functions depend only on their own contributions, audit outcomes, and public good provision, and do not involve beliefs about the common fund. 
By contrast, the Governor holds beliefs about what citizens expect to observe in the common fund, denoted by $\tau(\cdot)$, which are formed as functions of the citizens’ strategy profile. 
This modelling choice is consistent with standard practice in psychological game theory (see, e.g., \cite{BattigalliDufwenberg2009}, \cite{battigalli2022belief}). 
In particular, the Governor’s beliefs are strategy-based rather than outcome-based: they depend on the prescribed contribution probabilities and not on realized contribution outcomes.

%Belief-dependent preferences in the model are attributed exclusively to the Governor. 
%Citizens’ utility functions depend only on their own contributions, audit outcomes, and public good provision, and do not involve beliefs about the common fund. 
%By contrast, the Governor holds beliefs about what citizens expect to observe in the common fund, denoted by $\tau(\cdot)$, which are formed as functions of the citizens’ strategy profile. 
%In particular, beliefs are strategy-based rather than outcome-based: they depend on the prescribed contribution probabilities and not on realized contribution outcomes. 
%Consistent with standard practice in psychological game theory, when evaluating unilateral deviations from a candidate equilibrium, the Governor’s belief $\tau(\cdot)$ is held fixed at the level implied by the equilibrium strategy profile.

\paragraph{Interpretation of the reputational cost.}
The step-loss specification of the reputational cost should be understood as a tractable benchmark capturing norm-based or deontological concerns. In this interpretation, citizens (and hence the Governor) primarily care about whether embezzlement occurs at all, rather than about its precise magnitude: falling short of public expectations constitutes a salient norm violation that triggers a fixed reputational or psychological cost. This formulation is consistent with behavioural models of guilt aversion, reciprocity, and social norms, in which agents experience a discrete loss once a moral or social standard is violated, rather than a smoothly increasing disutility proportional to the size of the violation (see, e.g., 
\cite{BattigalliDufenberg2007},
\cite{BattigalliDufwenberg2009},
\cite{battigalli2022belief}). 

From an analytical perspective, the step-loss assumption allows us to obtain sharp thresholds and a complete characterization of equilibrium behavior, highlighting the interaction between audits, penalties, and reputational incentives. While alternative specifications with smoothly increasing reputational costs are plausible, they would preserve the same qualitative trade-offs while substantially complicating the analysis. We therefore adopt the step-loss function as a parsimonious benchmark that makes the underlying mechanisms transparent.

\section{Illustrative Example ($N=2,k=1$)} \label{sec:illustrative}

This section serves two purposes. First, it provides a fully worked-out example that makes transparent the interaction between audits, penalties, and reputational incentives. Second, it illustrates the equilibrium logic of the general model in the simplest non-trivial case, thereby clarifying the role of pivotality, belief formation, and the Governor’s threshold provision rule. All formal proofs of Observations 1–4 are provided in Appendix A.

%We illustrate the model in the simple case with two citizens,$N=2$, and one audit, $k=1$.
%Each citizen chooses a binary contribution $c_i\in\{0,1\}$, and the Governor observes only the realized common fund $X$. 

%\textcolor{blue}{(Here we say that exactly one citizen is audited, and previously we mention $k\in{1, 2,..,N}$; perhaps we should clarify this. I recall in our previous versions, we had upto $N-1$, or may be I am mistaken. I understand results, etc., do not change, but more clarificatory.)}

Figure 1 depicts the extensive form of this game. While the timing of the formal model specifies that contribution decisions precede the realization of the audit, the game tree represents Nature's move at the root. This transformation is strategically equivalent given the information structure, and allows us to clearly delineate the two possible ``states of the world'' regarding the audit target.

\paragraph{The Audit States.}
Nature moves first, selecting the state $\omega \in \{\omega_1, \omega_2\}$ with equal probability.

\begin{itemize}
    \item \textbf{State $\omega_1$ (Audit $C_1$):} In the left branch, Nature audits Citizen 1. Consequently, Citizen 1 is compelled to contribute regardless of their voluntary choice $c_1$. The total common fund $X$ is therefore determined solely by the voluntary contribution of the non-audited agent, Citizen 2. Specifically, $X = 1 + c_2$.

    \item \textbf{State $\omega_2$ (Audit $C_2$):} In the right branch, Nature audits Citizen 2. Here, Citizen 2 is compelled to contribute, and the total fund depends solely on Citizen 1, such that $X = c_1 + 1$.
\end{itemize}

\paragraph{Information Sets and Citizen Strategies.}
The dashed information sets $I_1$ and $I_2$ capture the imperfection of information. Citizen 1 (Red) chooses $c_1 \in \{0,1\}$ without observing the state of the world $\omega$. Citizen 2 (Blue) chooses $c_2 \in \{0,1\}$ observing neither $\omega$ nor $c_1$. This structure effectively models a simultaneous-move game between two citizens. Each citizen maximizes their expected utility, weighing the cost of contribution against the probability of being audited and the marginal benefit of the public good.

\paragraph{The Governor's Problem.}
The Governor (Green) moves last, selecting a provision level $g$. The Governor's feasible action set is constrained by the realized common fund, $g \in [0, X]$. Crucially, the Governor observes $X$ but not the specific contribution profile or audit state that generated it. This results in two distinct information sets for the Governor:

\begin{enumerate}
    \item \textbf{High Common Fund ($I_G^2$):} If the non-audited citizen contributes (i.e., $c_2=1$ in state $\omega_1$, or $c_1=1$ in state $\omega_2$), the common fund reaches its maximum size, $X=2$. The Governor chooses $g$ from the interval $[0, 2]$.

    \item \textbf{Low Common Fund ($I_G^1$):} If the non-audited citizen shirks (i.e., $c_2=0$ in state $\omega_1$, or $c_1=0$ in state $\omega_2$), the common fund consists only of the compelled contribution, $X=1$. The Governor is restricted to the interval $[0, 1]$.
\end{enumerate}

Since the Governor cannot distinguish between the histories leading to the same fund size, her strategy must map the aggregate fund $X$ directly to a provision level $g(X, \tau(p_1, p_2), \sigma)$.

\paragraph{Payoffs.}
%\textcolor{green}{In Figure 1 we only include terminal payoffs for only the scenario when Citizen 1 is audited. Since the payoffs are symmetrical we omit the case when Citizen 2 is audited.} 
As discussed in Section 2, the Governor's payoffs consist of two components: \textit{material payoff} and \textit{reputational payoff.} 
%Reputational payoff is determined by the difference between realized public good provision and the level that Governor believes citizens expect to observe. 

We make the following assumption about the Governor's beliefs.
Given that citizens contribute with probabilities $(p_1,p_2)$, Governor's belief about the expected common fund is:
\[
\tau(p_1,p_2) \;=\; \mathbb{E}[X] \;=\; 1 + \frac{p_1+p_2}{2}.
\]

%Now suppose that the Governor’s belief about the expected common fund is
%\[
%\tau(p_1,p_2) \;=\; \mathbb{E}[X] \;=\; 1 + \frac{p_1+p_2}{2},
%\]
%if citizens contribute with probabilities $(p_1,p_2)$.

%\textcolor{blue}{Perhaps rephrase to:
%Assuming citizens contribute with probabilities $(p_1,p_2)$, Governor's belief about the expected common fund is:
%\[
%\tau(p_1,p_2) \;=\; \mathbb{E}[X] \;=\; 1 + \frac{p_1+p_2}{2}.
%\]}

\begin{figure}[H]
\centering
\includegraphics[scale=0.7]{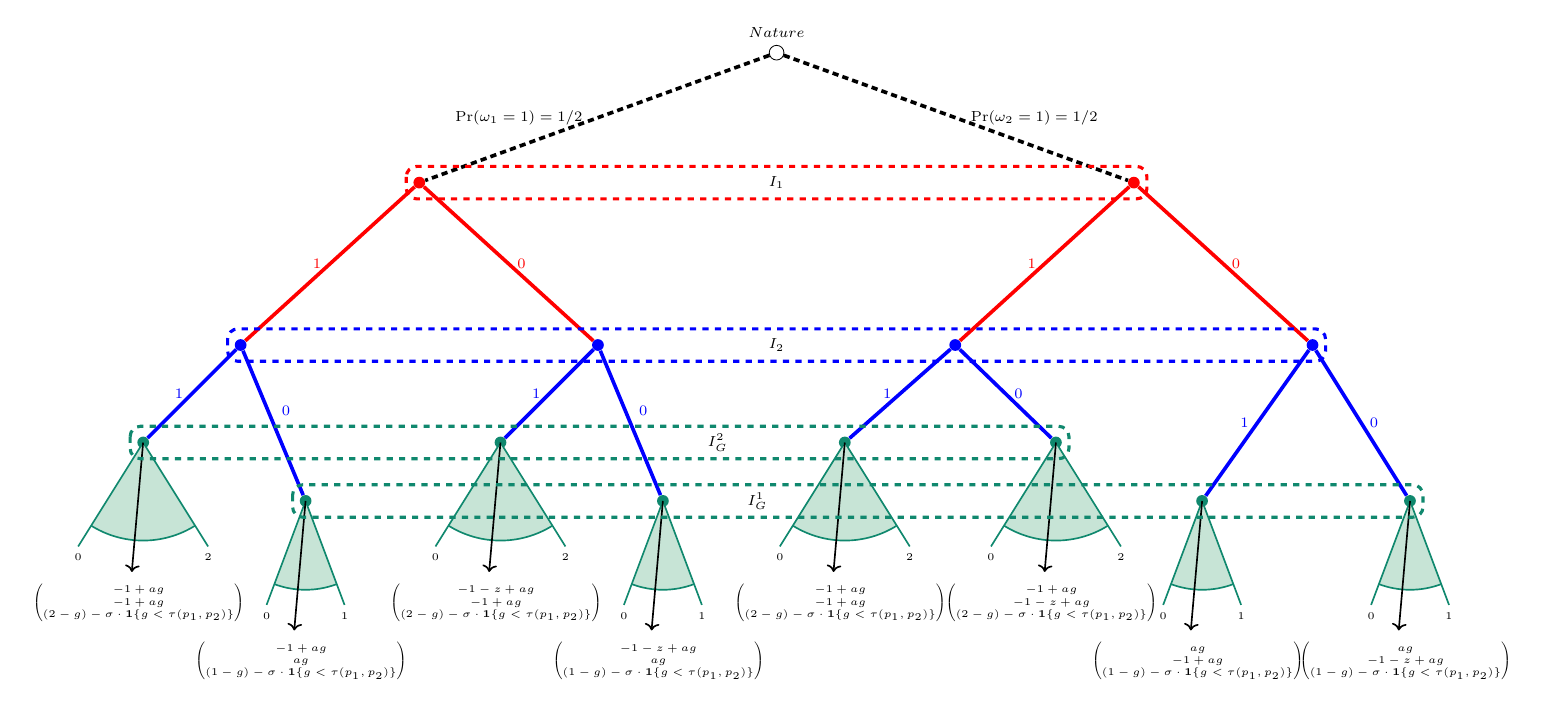}
\caption{Game tree for $N=2, k=1$.}
\label{fig:gametree}
\end{figure}

%The Governor provides $g=\tau(p_1,p_2)$ whenever $X\ge\tau(p_1,p_2)$ if she is reputation-sensitive ($\sigma\ge\tau$), and $g=0$ otherwise.
%\textcolor{blue}{Perhaps rephrase to: 

A reputation-sensitive Governor provides 

\[
g(X,\tau(p_1, p_2),\sigma) = \begin{cases}
    \tau(p_1,p_2), & X \ge \tau(p_1, p_2), \\[6pt]
    0, & X < \tau(p_1, p_2).
  \end{cases}
\]

The key insight of the example is that the Governor’s provision decision is driven by a threshold logic. Whenever the realized fund meets or exceeds the expected level $\tau(p_1, p_2)$, she prefers to meet expectations and avoid the reputational loss; otherwise, she fully embezzles the fund. This threshold structure generates discontinuous incentives for citizens: a unilateral deviation may not affect provision at all, or it may trigger a complete collapse of public good provision, depending on whether expectations are crossed.

%$g=\tau(p_1,p_2)$ whenever $X\ge\tau(p_1,p_2)$, and $g=0$ otherwise.

\subsubsection*{Free-Riding Equilibrium}

\begin{observation}[Free-Riding SPE]\label{prop:FR_N2}
Suppose $N=2$ and $k=1$.  
There exists a symmetric subgame perfect equilibrium in which both citizens shirk ($p_1 = p_2 =0$) if and only if
\[
z \le 1.
\]
In this equilibrium:
\begin{itemize}
\item the Governor provides $g=1$, if $\sigma \ge 1$;
\item the Governor provides $g=0$, if $\sigma \le 1$.
\end{itemize}
\end{observation}

\textit{Intuition.}  
When both citizens shirk, or $p_1 = p_2 = 0$, the expected fund is $\tau(0, 0) = 1$. A unilateral deviation does not affect public good provision but changes only the private cost. Shirking is optimal whenever the expected penalty from audits does not exceed the cost of voluntary contribution.

This equilibrium highlights an important feature of the model: when penalties are low, citizens’ contribution decisions are driven entirely by private cost considerations, since a unilateral deviation does not alter public good provision. The Governor’s reputational sensitivity affects only the level of provision conditional on universal shirking, but does not change individual incentives unless it is sufficiently strong to make deviations pivotal in other equilibria.

\subsubsection*{Full-Contribution Equilibrium}

\begin{observation}[Full-Contribution SPE]\label{prop:FC_N2}
Suppose $N=2$ and $k=1$.  
There exists a symmetric subgame perfect equilibrium in which both citizens contribute ($p_1 = p_2 =1$) if and only if one of the following conditions hold:
\begin{enumerate}
\item $\sigma \ge 2$ and $z \ge \max \{1 - 2a, 0\}$;
\item $\sigma < 2$ and $z \ge 1$.
\end{enumerate}
In case (1), the equilibrium is efficient and the Governor provides $g=2$.
In case (2), the Governor provides $g=0$ despite full contribution.
\end{observation}

\textit{Intuition.}  
With full contribution, each citizen is pivotal. If a citizen shirks and is not audited, the fund falls below expectations and a reputation-sensitive Governor provides no public good. This threat sustains efficient provision even for low penalties.

The full-contribution equilibrium illustrates how reputational incentives can substitute for formal punishment. When $\sigma$ is sufficiently large, each citizen becomes pivotal: a unilateral deviation may cause the realized fund to fall below expectations, triggering zero provision. Anticipating this discontinuity, citizens voluntarily contribute even when the penalty is low or zero, provided the marginal return from the public good is sufficiently high. This mechanism departs from standard public good models in which enforcement is purely monetary.

\subsubsection*{Asymmetric Profile}

\begin{observation}[Asymmetric SPE]\label{prop:asym_N2}
Suppose $N=2$. If $\sigma \ge 1.5$ and $z = 1 - 1.5a$, then there exists an asymmetric Subgame Perfect Equilibrium in which 
citizen~1 contributes ($c_1=1$), citizen~2 shirks ($c_2=0$),
the Governor’s belief about the expected size of the common fund is $\tau(1,0)=1.5$, and the Governor’s public good provision rule is:
\[
g(X,\tau(1,0),\sigma)=
\begin{cases}
0, & \text{if } X < \tau(1,0)=1.5,\\[6pt]
1.5, & \text{if } X \ge \tau(1,0)=1.5.\\[6pt]
\end{cases}
\]
\end{observation}

The Governor’s strategy depends on the realized common fund and on her belief $\tau(\cdot)$, which is determined by the citizens’ strategy profile and held fixed when evaluating deviations, consistent with standard psychological game theory.

This equilibrium exists only for the knife-edge parameter value $z = 1 - 1.5a$ and is therefore not robust to small perturbations of the parameters. It illustrates that asymmetric contributions can arise in equilibrium, even though such equilibria occur only on a restricted subset of the parameter space. This observation motivates our focus on symmetric equilibria in the general analysis.

The asymmetric equilibrium is sustained by finely balanced incentives: the contributing citizen is indifferent because her pivotal effect on expectations exactly offsets the private cost of contribution. Small perturbations in parameters break this balance, eliminating the equilibrium. For this reason, asymmetric equilibria are not robust and serve primarily as an illustration of how belief-dependent provision can generate heterogeneous contribution patterns.

%\begin{lemma}[Asymmetric Equilibrium]\label{prop:asym_N2}
%If $\sigma \ge 1.5$ and $z = 1 - 1.5a$,
%then the Governor’s belief is $\tau(1,0)=1.5$, and 
%$c_1 = 1$ and $c_2 = 0$ is a Nash equilibrium.
%\end{lemma}

%This equilibrium exists only for a particular parameter set $z = 1 - 1.5a$ and is therefore not robust. It serves to illustrate how asymmetric contribution behavior can arise.

\subsubsection*{Mixed-Strategy Equilibria}

%\textcolor{red}{Sakib will calculate this independelty- Done! It matches my calculation}

\begin{observation}[Mixed-Strategy SPE]\label{prop:mixed_N2}
Suppose $N=2$ and $k=1$.
\begin{enumerate}
\item If $\sigma \ge 2$, a unique symmetric mixed-strategy SPE exists with
\[
p^* = \frac{1-z}{a} - 1,
\]
provided $0 < a < 1$ and $\max \{1-2a, 0\} < z < 1-a$.
\item If $\sigma < 1$ and $z=1$, there exists a continuum of mixed-strategy SSPE with $g=0$.
\end{enumerate}
\end{observation}

Mixed-strategy equilibria arise when the marginal cost of contributing is exactly balanced by the expected marginal benefit from being pivotal. In the $N = 2$ example, this occurs only for a restricted parameter region. The example also shows that multiplicity may emerge: depending on parameters, there may exist either a unique mixed-strategy equilibrium or a continuum of equilibria. This preview anticipates the richer structure characterized in Section \ref{sec:mixed} for general $N$.

\subsubsection*{Summary}

\begin{figure}[h!]
    \centering
    \includegraphics[width=0.55\linewidth]{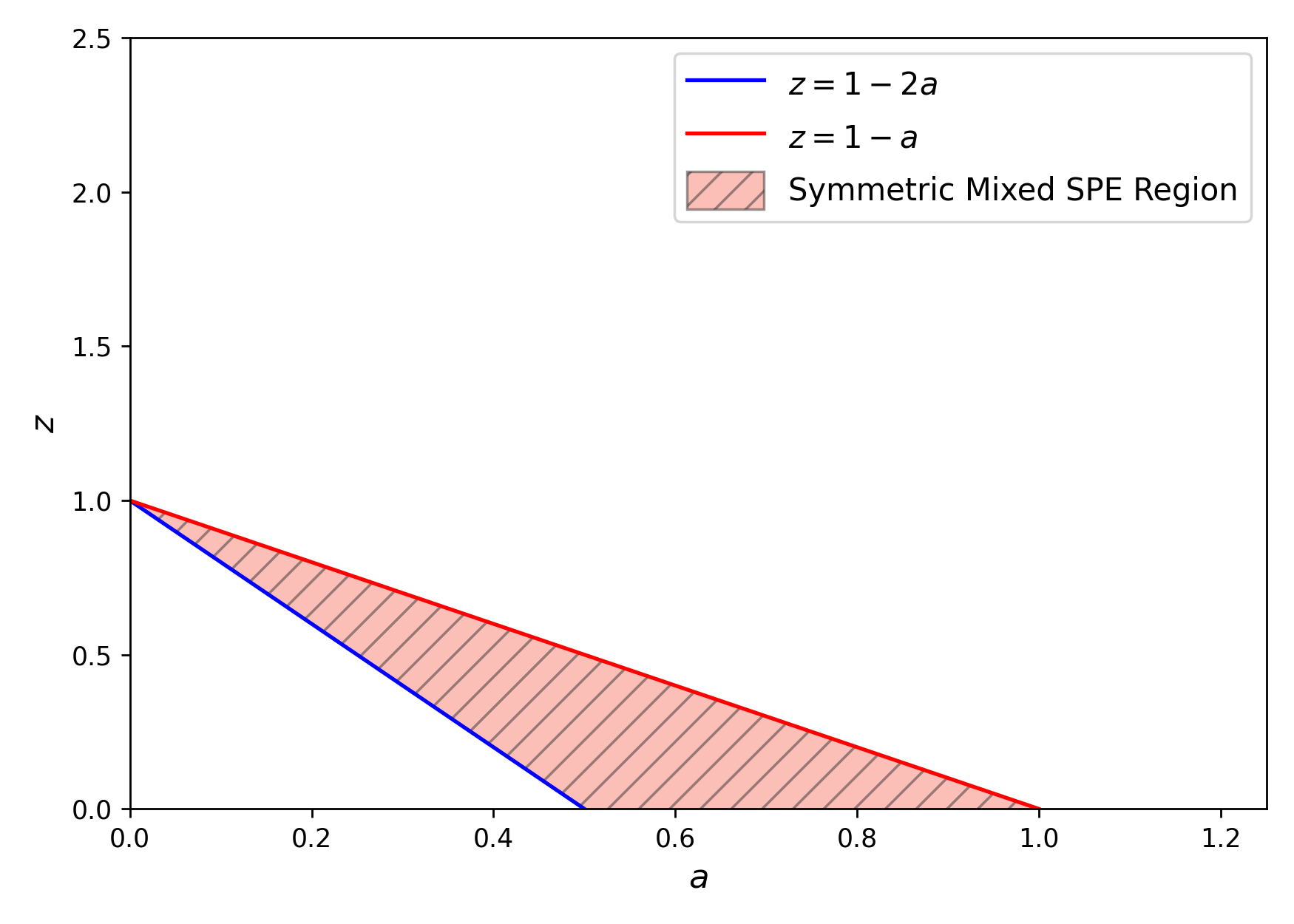}
        \caption{Feasible region for symmetric mixed-strategy SPE with $p \in (0,1)$, for $N=2$, $k=1$, $\sigma \geq 2$. The shaded region is bounded by $z = 1 - 2a$ (blue) and $z = 1 - a$ (red). A symmetric mixed-strategy SPE exists if and only if $(a,z)$ lies within this region, with equilibrium contribution probability $p^* = \frac{1-z}{a} - 1$.}

    \label{fig:p_N2K1}
\end{figure}

In summary, the $N = 2$ example illustrates all equilibrium phenomena present in the general model: free-riding equilibria, efficient full-contribution equilibria sustained by reputation, knife-edge asymmetric equilibria, and mixed-strategy equilibria. For $N=2$ and $k=1$, the model always admits a symmetric pure-strategy SPE.  
Figure \ref{fig:p_N2K1} shows that symmetric mixed-strategy  SPE exist for a particular parameter region. In that region, there may be either a unique or multiple symmetric mixed-strategy SPE.

A particular feature of this example is that the citizen who is not audited can infer the realized size of the common fund. Since exactly one citizen is audited in each period ($k=1$), a non-audited citizen knows that the other citizen must have been audited and therefore contributed. Consequently, she can deduce the total fund $X$ from her own action. This informational property is specific to the case $N=2$ with $k=1$ and simplifies the strategic structure of the example.
In the general model with $N>2$, when $k<N-1$, this feature disappears. A non-audited citizen cannot infer how many of the remaining citizens were audited or contributed voluntarily, and therefore does not know the realized size of the common fund. %The $N=2$ example thus provides a particularly transparent illustration of the mechanism, while the general analysis abstracts from this special informational structure.

Although the $N=2$ case provides maximal transparency, its informational structure is special. Appendix~B therefore examines the case $N=3$ with $k=1$, where individual citizens cannot infer the realized fund from their own audit status. The equilibrium logic remains qualitatively unchanged, demonstrating that the mechanisms identified in this section are not artifacts of the two-player setting.

\section{Analysis} \label{sec:analysis}
Asymmetric pure-strategy equilibria may arise in special cases, as shown in the illustrative example in Section 3, but they occur only on knife-edge regions of the parameter space. Given their lack of robustness and the symmetry of agents, information, and enforcement technology, we focus on symmetric equilibria, which exist generically and provide the natural benchmark for analysis.

For the rest of the paper, we focus on symmetric mixed-strategy profiles in which all citizens contribute with the same probability $p \in [0,1]$.  
Under symmetric voluntary contributions with probability $p$, the Governor believes citizens expect to observe the following common fund:

\[
\tau(p, ..., p) = \tau(p) = \mathbb{E}[X].
\]

We therefore focus on \textbf{Symmetric Subgame Perfect Equilibrium (SSPE)}.  
Citizens follow a symmetric strategy defined by a contribution probability $p \in [0,1]$.  
The Governor’s strategy is an allocation rule $g(X, \tau(p), \sigma)$ that assigns to each realized common fund $X$ a provision level $g \in [0,X]$.  
In the Symmetric Subgame Perfect Equilibrium, we follow standard practice in psychological game theory: the Governor’s beliefs are determined by the citizens’ strategy profile and are held fixed when evaluating unilateral deviations.

We will characterize two types of equilibria:  
\begin{itemize}
    \item \emph{Pure-strategy SSPE}, in which citizens choose deterministically, $p \in \{0,1\}$; and  
    \item \emph{Mixed-strategy SSPE}, in which citizens are indifferent and adopt probabilistic contributions with $p \in (0,1)$.  
\end{itemize}

We solve for the Symmetric Subgame Perfect Equilibria (SSPE) of the game by backward induction.  
The analysis proceeds in three steps. First, we derive 
the level of the common fund that the Governor believes citizens expect to observe, $\tau(p)$.  
Second, we characterize the Governor’s optimal allocation rule in Stage 3, given these expectations.  
Finally, taking the Governor’s allocation rule as given, we determine the citizens’ equilibrium contribution strategies in Stage 1.  

The results hinge on a central object of the model: the \emph{expected} common fund, $\tau(p)$.  
The following lemma characterizes these expectations.

\begin{lemma} \label{lemma1}
Suppose that there are $N$ citizens and $k$ of them are audited. If each citizen contributes 1 unit with probability $p$ at stage 1, then the expected number of units, $\tau(p)$, in the common fund is 
 \begin{align}
\tau(p) = k+p(N-k). \label{eq:1}
 \end{align}
  \end{lemma}

This result is intuitive: $k$ citizens are audited and therefore contribute to the common fund with certainty, while each of the $(N-k)$ non-audited citizens contribute with probability $p$.  
The formal proof of Lemma~\ref{lemma1} is provided in Appendix C.  

With the expected common fund formally defined, we can now state the Governor’s utility more explicitly. It depends on the realized common fund $X$, 
the citizens’ contribution probability $p$, which determines their expectations, and
her chosen public good provision level $g$, or 

\[
u_G(g,p;X,\sigma) = (X-g) - \sigma \cdot \mathbf{1}\{g < \tau(p)\}.
\]

The Governor’s problem is to choose an allocation rule that maximizes this utility for any given fund $X$, correctly anticipating citizens’ expectations.  
That is, for any $p \in [0,1]$, Lemma~\ref{lemma1} implies $\tau(p) = (N-k)p+k$, and the Governor solves
\[
\max_{g \in [0,X]} u_G(g,p;X,\sigma)
= \max_{g \in [0,X]} \big( (X-g) - \sigma \cdot \mathbf{1}\{g < (N-k)p + k\} \big).
\]

The following lemma characterizes the solution to this problem.

\begin{lemma}[Governor’s optimal rule]\label{Proposition1}
Given $p \in [0,1]$, the Governor’s optimal choice is
\[
g^{\ast}(X,\tau(p),\sigma) =
\begin{cases}
\tau(p), & \text{if } \min\{\sigma, X\} \geq \tau(p), \\[6pt]
0, & \text{otherwise.}
\end{cases}
\]
\end{lemma}
%{\textit{Proof of Lemma 2 is omitted.}
From {Lemma}~\ref{Proposition1}, it follows that if the Governor’s sensitivity is too low to meet expectations (i.e., $\sigma < \tau(p)$), then $g^{\ast}(X, \tau(p), \sigma) = 0$ for all $X \geq 0$.  

Given the Governor’s strategy in Stage~3, we now turn to the citizens’ problem in Stage~1 to characterize the Symmetric Subgame Perfect Equilibria.

\subsection{Pure-Strategy SSPE}

We now characterise the conditions under which all citizens adopt the same pure strategy.

\subsubsection{Free-Riding Equilibria ($p=0$).}

We begin by establishing the conditions under which a pure-strategy SSPE with universal shirking ($p=0$) exists.  
When $p=0$, the realized fund is $X=k$, and by Lemma~\ref{lemma1} the Governor’s expectation is $\tau(0)=k$.  
Accordingly, her optimal allocation rule is  
\[
    g^{\ast}(X,\tau(0),\sigma) =
    \begin{cases}
      k, & \text{if } \sigma \geq k, \\[6pt]
      0, & \text{if } \sigma < k.
    \end{cases}
\]

For this profile to constitute an equilibrium, no individual citizen should find it profitable to deviate.  
We verify this condition by considering two cases, depending on the Governor’s reputation sensitivity $\sigma$.

\paragraph{Case 1: Reputation-Sensitive Governor ($\sigma \ge k$).}

On the equilibrium path ($X=k$), the Governor provides $g^*(X, \tau(0), \sigma) = \tau(0) = k$.  
Suppose instead that a single citizen $i$ deviates by contributing. The common fund could then increase to $X' = k+1$, while the governor's expectation remains fixed at $\tau(0)=k$.  
Since the new common fund still meets expectations ($k+1 \geq k$), the Governor again provides $g^*(k+1,\tau(0),\sigma) = \tau(0) = k$.  

Thus, a deviation does not affect the provision level of the public good. 
In the free-riding SSPE, each citizen must weakly prefer shirking to contributing.  
The no-deviation condition is therefore  
\[
\mathbb{E}_{\omega_i}[u_i(0,\omega_i,k)] \;\geq\; \mathbb{E}_{\omega_i}[u_i(1,\omega_i,k)].
\]

The expected utility from shirking is  
\[
\mathbb{E}_{\omega_i}[u_i(0,\omega_i,k)] 
= \frac{k}{N}\,(-1-z+ak) + \frac{N-k}{N}\,(ak),
\]  
while the expected utility from contributing is  
\[
\mathbb{E}_{\omega_i}[u_i(1,\omega_i,k)] = -1 + ak.
\]

Substituting into the no-deviation condition gives  
\[
\frac{k}{N}(-1-z+ak) + \frac{N-k}{N}(ak) \;\geq\; -1 + ak.
\]

This condition reduces to   
\[
z \leq \frac{N-k}{k}.
\]

\paragraph{Case 2: Reputation-Insensitive Governor ($\sigma < k$).}

On the equilibrium path ($X=k$), the Governor provides $g^*(X, \tau(0), \sigma) = 0$.  
If a single citizen $i$ deviates by contributing, the common fund could then increase to $X' = k+1$, but the Governor’s belief remains $\tau(0) = k$, and  
she again chooses $g^*(k+1,\tau(0),\sigma) = 0$.  

As in the previous case, a deviation does not affect the public good provision (which remains zero).  
Hence, the no-deviation condition is  
\[
\mathbb{E}_{\omega_i}[u_i(0, \omega_i, 0)] \;\ge\; \mathbb{E}_{\omega_i}[u_i(1, \omega_i, 0)],
\]  
which simplifies to  
\[
z \le \frac{N-k}{k}.
\]

We can now state our first result.  

\begin{proposition}[Free-Riding]\label{Theorem:1}
If the penalty $z$ satisfies
\[
0 \leq z \leq \frac{N-k}{k},
\]
then for any marginal per capita return $a \in (0,1)$,
the following strategy profile:
\begin{enumerate}[label=(\roman*)]
    \item each citizen shirks ($c_i = 0$ for all $i \in \mathcal{I}$),
    \item the Governor follows the allocation rule
    \[
    g^{\ast}(X,\tau(0);\sigma) =
    \begin{cases}
        k, & \text{if } \sigma \geq k, \\[6pt]
        0, & \text{if } \sigma < k,
    \end{cases}
    \]
\end{enumerate}
constitutes a Pure-strategy Symmetric Subgame Perfect Equilibrium. 
\end{proposition}
%{\textit{Proof of Proposition 1 is omitted.}

\subsubsection{Full-Contribution Equilibria ($p=1$).}

We now establish the conditions under which a pure-strategy SSPE with universal contribution ($p=1$) exists.  
When $p=1$, the realized fund is $X=N$, and by Lemma~\ref{lemma1} the Governor’s expectation is $\tau(1)=N$.  
Accordingly, her optimal allocation rule is  
\[
    g^{\ast}(X,\tau(1),\sigma) =
    \begin{cases}
      N, & \text{if } \min\{\sigma, X\} \geq N, \\[6pt]
      0, & \text{if } \min\{\sigma, X\} < N.
    \end{cases}
\]

For this profile to constitute an equilibrium, no citizen should have an incentive to deviate by shirking.  
We verify this condition by considering two cases, depending on the Governor’s reputation sensitivity $\sigma$.

\paragraph{Case 1: Reputation-Sensitive Governor ($\sigma \ge N$).}

On the equilibrium path ($X=N$), the Governor provides $g^*(N,\tau(1),\sigma) = \tau(1) = N$.  
Consider a citizen $i$ who deviates by shirking. The resulting fund $X'$ depends on whether this citizen is audited:
\begin{itemize}
    \item If citizen $i$ is not audited (probability $\tfrac{N-k}{N}$), the fund becomes $X' = N-1$. Since $X' < \tau(1)$, the Governor provides $g=0$.
    \item If citizen $i$ is audited (probability $\tfrac{k}{N}$), the contribution is enforced, so the fund is $X' = N$. Since $X' \geq \tau(1)$, the Governor provides $g=N$.
\end{itemize}

For the full-contribution profile to be an equilibrium, a citizen must weakly prefer contributing to shirking.  
The no-deviation condition is therefore
\[
\mathbb{E}_{\omega_i}[u_i(1, \omega_i, N)] \;\geq\; \mathbb{E}_{\omega_i}[u_i(0, \omega_i, g')].
\]

The utility from contributing is 

\[
\mathbb{E}_{\omega_i}[u_i(1, \omega_i, N)] = -1 + aN.
\]

The expected utility from shirking is
\[
\mathbb{E}_{\omega_i}[u_i(0, \omega_i, g')] 
= \frac{k}{N}\,(-1 - z + aN) + \frac{N-k}{N}\,(0).
\]

Substituting into the no-deviation condition yields
the condition on the penalty
\[
z \;\geq\; \frac{N-k}{k}(1 - aN).
\]

\paragraph{Case 2: Reputation-Insensitive Governor ($\sigma < N$).}

On the equilibrium path ($X=N$), the Governor provides $g^*(N,\tau(1),\sigma) = 0$.  
If a citizen deviates, the fund becomes either $X' = N-1$ or $X' = N$.  
In both cases, since the Governor is reputation-insensitive, she provides $g=0$.  

Thus, a deviation has no effect on the level of the public good (which remains zero).  
The no-deviation condition is  
\[
\mathbb{E}_{\omega_i}[u_i(1, \omega_i, 0)] \;\geq\; \mathbb{E}_{\omega_i}[u_i(0, \omega_i, 0)],
\]  
which simplifies to  
\[
z \;\geq\; \frac{N-k}{k}.
\]

We can now state our second result.

\begin{proposition}[Full-Contribution]\label{Theorem:2}
\begin{enumerate}[label=(\alph*)]
    \item \textit{Reputation-sensitive Governor}: Suppose that $\sigma \geq N$.  
    Consider the strategy profile where each citizen contributes ($c_i=1$) and the Governor follows the allocation rule
    \[
    g^*(X,\tau(1),\sigma) = 
    \begin{cases}
        N, & \text{if } X \geq N, \\[6pt]
        0, & \text{otherwise}.
    \end{cases}
    \]  
    This profile is a Pure-strategy Symmetric Subgame Perfect Equilibrium if   
    \[
    z \;\geq\; \frac{N-k}{k}(1-aN).
    \]
    \item \textit{Reputation-insensitive Governor}: Suppose that $\sigma < N$.  
    Consider the strategy profile where each citizen contributes ($c_i=1$) and the Governor plays the allocation rule $g^*(X,\tau(1),\sigma) = 0$.  
    This profile is a Pure-strategy Symmetric Subgame Perfect Equilibrium if   
    \[
    z \;\geq\; \frac{N-k}{k}.
    \]
\end{enumerate}
\end{proposition}
%\textcolor{red}{\textit{Proof of Proposition 2 is omitted.}}

\begin{figure}[h!]
    \centering
    \includegraphics[width=0.99\linewidth]{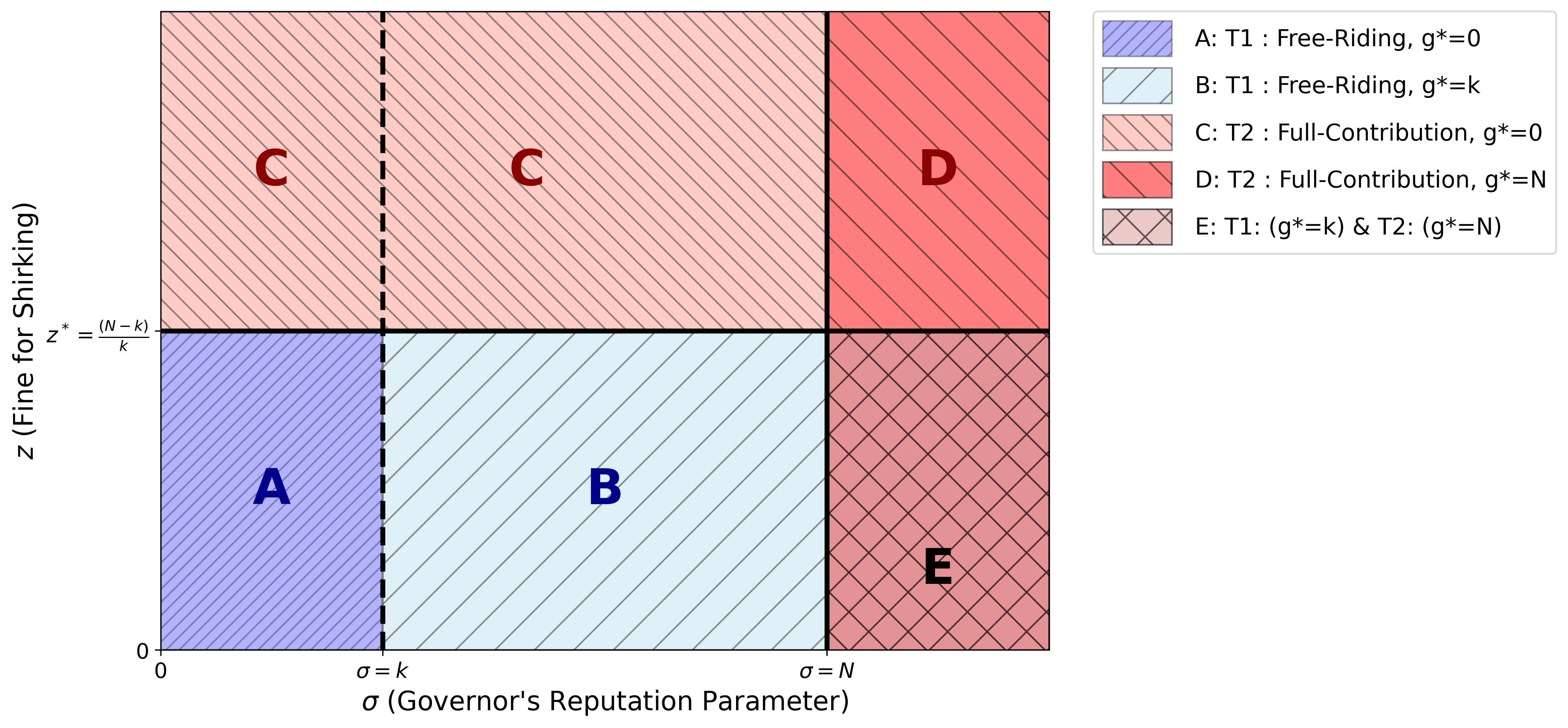}
    \caption{Visual characterisation of the pure-strategy equilibria in the \((\sigma, z)\) parameter space.}
    \label{fig:thm1_thm2_comp}
\end{figure}

Figure~\ref{fig:thm1_thm2_comp} provides a complete visual characterization of the pure-strategy SSPE in the $(\sigma, z)$ parameter space, with five distinct regions labeled A through E.  
The analysis shows that the penalty for shirking, $z$, is the primary determinant of citizen behavior.  
The critical boundary is the horizontal line at $z^* = (N-k)/k$.

%\textcolor{red}{Discuss efficiency. \newline
%Discuss the intersection of both theorem 1 and 2. \newline
%Make sure the statements in the figure matches the statement in the paper. Change the labels in the figure}

For penalties below this threshold ($z < z^*$), the conditions of Proposition~\ref{Theorem:1} apply, and the pure-strategy SSPE are \textit{Free-Riding} SSPE.  
Within this low-penalty region, the Governor's reputation sensitivity $\sigma$ plays a secondary role.  
In \textbf{Region~A} ($\sigma < k$), the Governor is reputation-insensitive: all citizens free-ride and no public good is provided ($g^*=0$).  
In \textbf{Region~B} ($k \leq \sigma < N$), the Governor is sufficiently reputation-sensitive to meet the low expectations generated by universal shirking: all citizens still free-ride, but the Governor provides a minimal amount equal to the audited contributions ($g^*=k$).

For penalties above the critical threshold ($z > z^*$), the conditions of Proposition~\ref{Theorem:2} hold, and the pure-strategy SSPE are \textit{Full Contribution} SSPE.  
Here again, the Governor's reputation sensitivity determines the provision outcome.  
In \textbf{Region~C} ($\sigma < N$), full contribution is enforced by the high penalty, but the Governor is not sufficiently reputation-sensitive to provide the public good, so the entire fund is embezzled ($g^*=0$).  
In \textbf{Region~D} ($\sigma \geq N$), the Governor is highly reputation-sensitive: full contribution is sustained and the Governor provides the full amount ($g^*=N$).

An important implication of Proposition~\ref{Theorem:2} concerns the penalty requirement in Region~D.  
When the Governor is reputation-sensitive ($\sigma \geq N$), the condition for full contribution is $z \geq \frac{N-k}{k}(1 - aN)$.  
Note that $1 - aN$ is \textit{not} assumed to be negative; its sign depends on parameters.  
When the efficiency condition holds ($aN > 1$), the right-hand side is negative and the condition is automatically satisfied for any $z \geq 0$.  
In this case, there exists an efficient full-contribution pure-strategy SSPE in which all citizens contribute and the Governor provides the full amount $g^*=N$, even when the penalty is zero.  
This surprising result departs from standard public good models and arises because, in this SSPE, each citizen is pivotal: if any one citizen shirks and is not audited, the Governor will no longer have sufficient funds to meet expectations and is therefore forced to provide zero public goods.  
Anticipating this, every citizen chooses to contribute, and the efficient outcome is sustained.  
When instead $aN \leq 1$, the right-hand side is positive and a genuinely binding penalty ($z > 0$) is required to deter shirking, since the public good return alone is insufficient to make each citizen pivotal.

Finally, \textbf{Region~E} ($\sigma \geq N$, $z < z^*$) identifies a region where the conditions of both Proposition~\ref{Theorem:1} and Proposition~\ref{Theorem:2} are satisfied simultaneously: the Governor is reputation-sensitive, but the penalty is low.  
In this region, the game admits two coexisting types of SSPE: a free-riding equilibrium with $g^*=k$, and an efficient full-contribution equilibrium with $g^*=N$.  
The coexistence of these equilibria underscores the importance of coordination and expectation management when reputational incentives are strong but formal penalties are weak.

%The figure also reveals the possible of mixed strategy equilibrium. 

\section{Mixed-Strategy SSPE} \label{sec:mixed}

In this section, we analyze mixed-strategy SSPE, in which each citizen must be indifferent between contributing and shirking.  
We distinguish two cases according to the Governor’s reputation sensitivity:  
(i) $\sigma < \tau(p)$, and  
(ii) $\sigma \geq \tau(p)$.

\subsection{Case 1: Reputation-Insensitive Governor, $\sigma < \tau(p)$.}

When the Governor is reputation-insensitive, $\sigma < \tau(p)$, Lemma~\ref{Proposition1} implies that her optimal strategy is always to provide no public good, $g^*(X, \tau(p), \sigma) = 0$, regardless of the realized common fund $X$.  
Since the public good is zero, the indifference condition reduces to
\[
\mathbb{E}_{\omega_i}[u_i(0, \omega_i, 0)] 
= \mathbb{E}_{\omega_i}[u_i(1, \omega_i, 0)] 
\quad \Longleftrightarrow \quad 
-\tfrac{k}{N}(1+z) = -1,
\]
which yields
\[
z = \frac{N-k}{k}.
\]

Hence, we obtain the following result.

\begin{proposition}[Continuum of mixed equilibria]\label{Theorem:3_continuum}
Suppose $z = \tfrac{N-k}{k}$.  
Then, for any $a \in (0,1)$ and any $p \in (0,1)$ such that $(N-k)p + k > \sigma$, the strategy profile in which citizens contribute with probability $p$ and the Governor provides $g=0$ constitutes a Mixed-strategy Symmetric Subgame Perfect Equilibrium.

\end{proposition}

Proposition~\ref{Theorem:3_continuum} shows that for a particular penalty there is a continuum of mixed-strategy SSPE.  
The following corollary highlights an important special case.

\begin{corollary}\label{Corollary:continuum}
If $\sigma < k$ and $z = \tfrac{N-k}{k}$, then for any $a \in (0,1)$ and any $p \in (0,1)$, a strategy profile where citizens contribute with probability $p$ and the Governor provides $g=0$ constitutes a Mixed-strategy Symmetric Subgame Perfect Equilibrium.
\end{corollary}

This corollary emphasizes that, when the Governor is reputation-insensitive, $\sigma < k$, any interior contribution probability can be sustained in equilibrium, even though no public good is provided.  
It illustrates why the equilibrium set may be very large in this regime, despite the inefficiency of the outcome.

\subsection{Case 2: Reputation-Sensitive Governor, $\sigma \ge \tau(p)$.}

When the Governor is reputation-sensitive, $\sigma \geq \tau(p)$, Lemma~\ref{Proposition1} implies that she provides the public good whenever the common fund is sufficient:  
\[
    g^{\ast}(X, \tau(p), \sigma) =
    \begin{cases}
      \tau(p), & \text{if } X \geq \tau(p), \\[6pt]
      0, & \text{if } X < \tau(p).
    \end{cases}
\]

In a mixed-strategy SSPE, each citizen must be indifferent between contributing and shirking.  
For citizen $i$, the indifference condition is
\[
\mathbb{E}_{\omega_i}\!\left[u_i(0,\omega_i,g^{\ast}(\omega_i+X_{-i},\tau(p),\sigma))\right]
= \mathbb{E}_{\omega_i}\!\left[u_i(1,\omega_i,g^{\ast}(1+X_{-i},\tau(p),\sigma))\right],
\]
where
\[
X = \sum_{j=1}^N \max\{c_j,\omega_j\} 
= \max\{c_i,\omega_i\} + X_{-i}.
\]

Expanding expectations over the audit outcome gives,

\begin{align*}
&\frac{k}{N}\, u_i(0,1,g^{\ast}(1+X_{-i},\tau(p),\sigma))
+ \frac{N-k}{N}\, u_i(0,0,g^{\ast}(X_{-i},\tau(p),\sigma)) \\
&= \frac{k}{N}\, u_i(1,1,g^{\ast}(1+X_{-i},\tau(p),\sigma)) 
+ \frac{N-k}{N}\, u_i(1,0,g^{\ast}(1+X_{-i},\tau(p),\sigma)).
\end{align*}

If citizen $i$ is audited ($\omega_i=1$), her initial choice $c_i$ does not affect the realized fund $(1+X_{-i})$, and the only difference in utility is the penalty.  
Substituting and simplifying yields,
\[
k(-1-z + a\,g^{\ast}(1+X_{-i},\tau(p),\sigma)) 
+ (N-k)a\,g^{\ast}(X_{-i},\tau(p),\sigma) 
= N(-1 + a\,g^{\ast}(1+X_{-i},\tau(p),\sigma)).
\]

Rearranging the above expression, gives the indifference condition,

\[
1 - \frac{k}{N-k}z
= a\left(g^{\ast}(1+X_{-i},\tau(p),\sigma) - g^{\ast}(X_{-i},\tau(p),\sigma)\right),
\]
where, the left-hand side represents the \emph{marginal cost} of contributing, while the right-hand side captures the \emph{marginal benefit}.  
A mixed-strategy SSPE exists whenever this condition admits a solution $p^* \in (0,1)$.

To analyse the existence of mixed-strategy equilibria, it is convenient to define a \emph{Net Gain function}. For given parameters $(a,z)$, the Net Gain function $f(p)$ measures the difference between the expected marginal benefit from contributing and the marginal cost of doing so when all other citizens contribute with probability $p$. Formally,
\[
f(p) \;=\; a \cdot \mathbb{E}_{c_{-i}}\!\left[g^*((1 + X_{-i}), \tau(p), \sigma) - g^*(X_{-i}, \tau(p), \sigma)\right] \;-\; \left(1 - \frac{k z}{N-k}\right).
\]

A positive value of $f(p)$ means that contributing yields a higher expected payoff than shirking, while a negative value implies the opposite.  

\medskip
The public good benefit arises only when the citizen’s contribution is \emph{pivotal}.  
Let $m(p) = \lceil \tau(p) \rceil$.  
A citizen is pivotal when exactly $m(p)-1-k$ of the remaining $N-1-k$ non-audited citizens contribute voluntarily.  
The probability of this pivotal event is given by the binomial probability
\[
\Pr(\text{pivotal}) \;=\; \binom{N-1-k}{\,m(p)-1-k\,}\, p^{\,m(p)-1-k} (1-p)^{\,N-m(p)}.
\]

When the citizen is pivotal, the Governor sets the public good provision equal to $\tau(p)$.  
Combining these observations gives the closed-form expression for the Net Gain function:
\begin{equation}\label{eq:fp_expanded}
    f(p) \;=\; a \cdot \tau(p) \cdot 
    \binom{N-1-k}{\,m(p)-1-k\,} \,
    p^{\,m(p)-1-k} (1-p)^{\,N-m(p)}
    \;-\; \left(1 - \frac{k z}{N-k}\right).
\end{equation}

\medskip
Expression~\eqref{eq:fp_expanded} shows that although the formula for $f(p)$ changes whenever $\tau(p)$ crosses an integer, the function itself is continuous on $(0,1)$.  
At these integer thresholds $f(p)$ exhibits \emph{kinks}, but no discontinuities, as established in the following lemma.

\begin{lemma}[Continuity of the Net Gain Function]\label{lemma:continuity}
The Net Gain function $f(p)$ is continuous for all $p \in (0,1)$.
\end{lemma}

The formal proof of Lemma~\ref{lemma:continuity} is in Appendix C. Figure~\ref{fig:my_grid_fp} illustrates the shape of the Net Gain function $f(p)$ for several parameter values. The plots visually confirm the structure established in Lemma~\ref{lemma:continuity}: the function consists of distinct polynomial segments joined at kinks. The vertical red lines in Figure \ref{fig:my_grid_fp} shows these distinct polynomial segments.  The equilibria occurs when the function crosses the x axis. 
%(blue curve). 

The intuition behind the continuity proof is twofold. First, between any two kinks the integer threshold for public good provision is constant, so the pivotal probability reduces to a single polynomial in $p$. Second, at each kink point where the threshold changes, the formula for the pivotal probability also changes in exactly the right way to ensure that the two adjacent polynomial pieces meet smoothly, with left- and right-hand limits equal.

 Intuitively, the Net Gain function $f(p)$ shows us the trade-off between the net marginal cost of contributing (adjusted for the avoided penalty) and the expected marginal benefit of triggering public good provision. Crucially, this benefit materializes only when a citizen is pivotal: when the aggregate contribution from others falls exactly one unit short of the Governor's reputational threshold, $\tau(p)$. Although the Governor's provision rule is discrete, jumping from zero to full provision once the threshold is met, the uncertainty regarding audit outcomes and the actions of other citizens smooths the incentives. The kinks correspond to critical points where citizens' expectations cross an integer value, effectively raising the threshold for provision. Continuity ensures that these critical points do not result in abrupt discontinuities in expected payoffs, allowing citizens to calibrate their contribution probability to balance the expected penalty from shirking against the potential benefit from enabling public good provision.

Figure \ref{fig:my_grid_fp}  also reveals that $f(p)$ may cross the $x$-axis multiple times, implying the possibility of multiple mixed-strategy equilibria. This observation motivates the existence result in Proposition~\ref{Thoerem: 4 Existence>N}. The underlying intuition is based on the Intermediate Value Theorem. Rather than solving for the roots explicitly, we derive conditions on the parameters $(a,z)$ that force the continuous function $f(p)$ to begin below the $x$-axis (i.e., $f(0^{+}) < 0$) and end above it (i.e., $f(1^{-}) > 0$). By continuity, a function that starts negative and ends positive must cross the axis at least once, thereby guaranteeing the existence of a mixed-strategy SSPE.

%The plots also reveal that $f(p)$ may cross the $x$-axis multiple times, implying the possibility of multiple mixed-strategy equilibria.  
%This observation motivates the existence result in Proposition~\ref{Thoerem: 4 Existence>N}.  

\begin{figure}[ht!]
    \centering
    \includegraphics[width=0.90\linewidth]{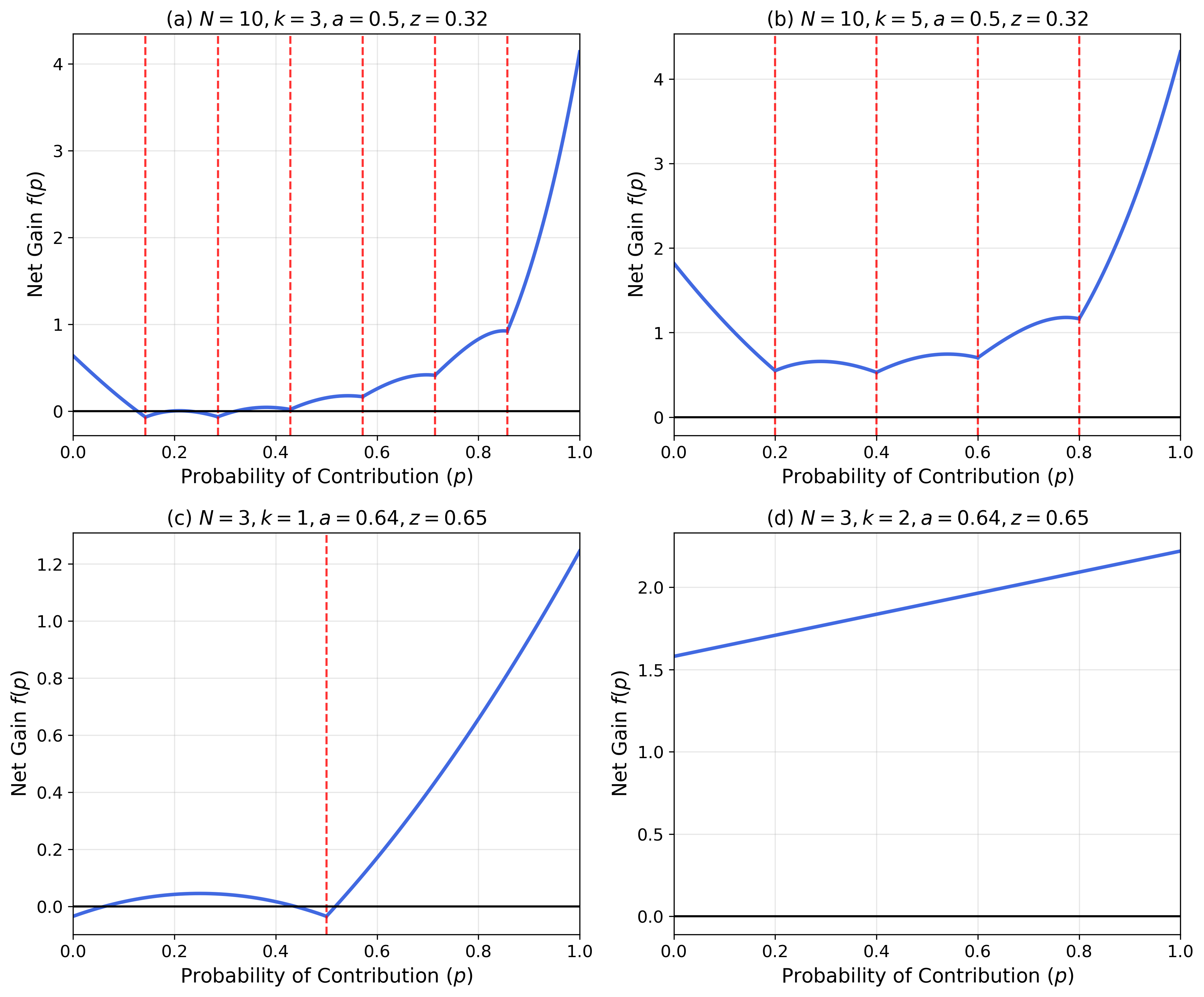}
    \caption{Plot of the Net Gain function $f(p)$ for different parameter values. The blue curve shows $f(p)$, which measures the difference between the expected marginal benefit of contributing and the marginal cost. Mixed-strategy SSPE correspond to values of $p$ where the blue curve crosses the $x$-axis, i.e., $f(p)=0$. The vertical red dashed lines mark the kink points where citizens' expectations $\tau(p)$ cross an integer, delineating the distinct polynomial segments of $f(p)$. Panels~(a) and~(c) exhibit multiple mixed-strategy SSPE, while panels~(b) and~(d) show cases where $f(p)>0$ for all $p \in (0,1)$, implying no mixed-strategy SSPE exists.}
    \label{fig:my_grid_fp}
\end{figure}

\newpage
\begin{proposition}[Existence of Mixed-Strategy Equilibria]\label{Thoerem: 4 Existence>N}
  If the following conditions hold
   \begin{enumerate}[label=(\roman*)]
       \item  $0<a<1 / k$,
       \item $\max \left\{0, \frac{N-k}{k}(1-a N)\right\} \leq z<\frac{N-k}{k}(1-a k)$,
       \item $\sigma \geq N$,
   \end{enumerate}
   then there exists at least one Mixed-strategy Symmetric Subgame Perfect Equilibrium.
\end{proposition}

The formal proof of Proposition \ref{Thoerem: 4 Existence>N} is in Appendix C.

\section{Conclusion} \label{sec:conclusion}

This paper develops and analyzes a model of public good provision with audits, penalties, and a self-interested Governor who may embezzle funds but also cares about citizens’ expectations. 
We provide a complete characterization of \emph{symmetric} pure-strategy SPE and of \emph{symmetric} mixed-strategy SPE.
Our results highlight how the interaction between formal penalties and reputational incentives shapes equilibrium behavior. In particular, we show that the model always admits at least one pure-strategy SSPE, ranging from full free-riding with complete embezzlement to efficient public good provision with no embezzlement, depending on the Governor’s reputation sensitivity and the level of penalties. Mixed-strategy SSPE exist within a narrow region of parameters.

Beyond the theoretical contributions, our model connects with the broader literature on political corruption, embezzlement, and psychological game theory. It shows how belief-dependent preference, such as guilt aversion or reciprocity, can serve as a disciplining device in taxation and public good environments. These findings suggest policy-relevant insights: while penalties and audits are important, reputational concerns can play an equally crucial role in sustaining efficient outcomes. At the same time, the coexistence of efficient and free-riding equilibria underscores the importance of coordination and expectation management. Future work could extend the analysis to asymmetric strategies, heterogeneous populations, or dynamic settings, providing further insight into the interplay between formal enforcement and reputational mechanisms in public good provision.

\printbibliography

@article{gallice2019co,
  title={Co-operation in Social Dilemmas Through Position Uncertainty},
  author={Gallice, Andrea and Monz{\'o}n, Ignacio},
  journal={The Economic Journal},
  volume={129},
  number={621},
  pages={2137--2154},
  year={2019},
  publisher={Oxford University Press}
}

@article{van2013candidates,
  title={Candidates, credibility, and re-election incentives},
  author={Van Weelden, Richard},
  journal={Review of Economic Studies},
  volume={80},
  number={4},
  pages={1622--1651},
  year={2013},
  publisher={Oxford University Press}
}

@article{fox2015social,
  title={Social accountability: what does the evidence really say?},
  author={Fox, Jonathan A},
  journal={World Development},
  volume={72},
  pages={346--361},
  year={2015},
  publisher={Elsevier}
}

@article{lambert2015social,
  title={Social accountability to contain corruption},
  author={Lambert-Mogiliansky, Ariane},
  journal={Journal of Development Economics},
  volume={116},
  pages={158--168},
  year={2015},
  publisher={Elsevier}
}

@article{alm2019motivates,
  title={What Motivates Tax Compliance?},
  author={Alm, James},
  journal={Journal of Economic Surveys},
  volume={33},
  number={2},
  pages={353--388},
  year={2019},
  publisher={Wiley Online Library}
}

@article{besley2023norms,
  title={Norms, enforcement, and tax evasion},
  author={Besley, Timothy and Jensen, Anders and Persson, Torsten},
  journal={Review of Economics and Statistics},
  volume={105},
  number={4},
  pages={998--1007},
  year={2023},
  publisher={MIT Press One Rogers Street, Cambridge, MA 02142-1209, USA journals-info~…}
}

@article{Markussen2016,
author = {Markussen, Thomas and Putterman, Louis and Tyran, Jean-Robert},
journal = {European Economic Review},
month = Oct,
pages = {372--388},
title = {{Judicial error and cooperation}},
volume = {89},
year = {2016}
}

@article{Markussen2014,
author = {Markussen, Thomas and Putterman, Louis and Tyran, Jean-Robert},
journal = {Review of Economic Studies},
number = {1},
pages = {301--324},
title = {{Self-organization for collective action: An experimental study of voting on sanction regimes}},
volume = {81},
year = {2014}
}

@article{Baldassarri2012,
author = {Baldassarri, Delia and Grossman, Guy},
journal = {American Journal of Political Science},
month = {10},
number = {4},
pages = {964 - 985},
pmid = {21690401},
publisher = {Wiley},
title = {{The impact of elections on cooperation: evidence from lab-in-the-field experiment in Uganda}},
volume = {56},
year = {2012}
}

@article{Chaudhuri2011,
author = {Chaudhuri, Ananish},
journal = {Experimental Economics},
month = {03},
number = {1},
pages = {47--83},
publisher = {Springer US},
title = {{Sustaining cooperation in laboratory public goods experiments: A selective survey of the literature}},
volume = {14},
year = {2011}
}

@article{fehr2000Gachter,
  title={Cooperation and Punishment in Public Goods Experiments},
  author={Fehr, Ernst and G{\"a}chter, Simon},
  journal={American Economic Review},
  volume={90},
  number={4},
  pages={980--994},
  year={2000}
}

@article{allingham1972income,
  title={Income tax evasion: A theoretical analysis},
  author={Allingham, Michael G and Sandmo, Agnar},
  journal={Journal of Public Economics},
  volume={1},
  number={3-4},
  pages={323--338},
  year={1972},
  publisher={Elsevier}
}

@article{welch1997effects,
  title={The effects of charges of corruption on voting behavior in congressional elections, 1982-1990},
  author={Welch, Susan and Hibbing, John R},
  journal={Journal of Politics},
  volume={59},
  number={1},
  pages={226--239},
  year={1997},
  publisher={University of Texas Press}
}

@incollection{ledyard95, 
title={Public Goods: A Survey of Experimental Research}, 
booktitle={The {H}andbook of {E}xperimental {E}conomics}, publisher={Princeton University Press}, author={Ledyard, John}, editor={Kagel , John H and Roth, Alvin}, year={1995}}

@article{Litina2016,
author = {Litina, Anastasia and Palivos, Theodore},
journal = {Journal of Economic Behavior and Organization},
pages = {164--177},
publisher = {Elsevier B.V.},
title = {{Corruption, tax evasion and social values}},
volume = {124},
year = {2016}
}

@article{Reinikka2004,
author = {Reinikka, Ritva and Svensson, Jakob},
title = {Local Capture: Evidence from a Central Government Transfer Program in {U}ganda},
journal = {Quarterly Journal of Economics},
volume = {119},
number = {2},
pages = {679-705},
year = {2004},
}

@article{fehr2002altruistic,
  title={Altruistic punishment in humans},
  author={Fehr, Ernst and G{\"a}chter, Simon},
  journal={Nature},
  volume={415},
  number={6868},
  pages={137},
  year={2002},
  publisher={Nature Publishing Group}
}

@article{ferraz2011electoral,
  title={Electoral accountability and corruption: Evidence from the audits of local governments},
  author={Ferraz, Claudio and Finan, Frederico},
  journal={American Economic Review},
  volume={101},
  number={4},
  pages={1274--1311},
  year={2011}
}

@article{andreoni1996governmen,
 title={Do government subsidies increase the private supply of public goods?},
  author={Andreoni, James and Bergstrom, Ted},
  journal={Public Choice},
  volume={88},
  number={3-4},
  pages={295--308},
  year={1996},
  publisher={Springer}
}

@article{ades1999rents,
  title={Rents, competition, and corruption},
  author={Ades, Alberto and Di Tella, Rafael},
  journal={American Economic Review},
  volume={89},
  number={4},
  pages={982--993},
  year={1999}
}

@article{brollo2013political,
  title={The political resource curse},
  author={Brollo, Fernanda and Nannicini, Tommaso and Perotti, Roberto and Tabellini, Guido},
  journal={American Economic Review},
  volume={103},
  number={5},
  pages={1759--96},
  year={2013}
}

@article{weitz2017can,
  title={Can citizens discern? {Information} credibility, political sophistication, and the punishment of corruption in {B}razil},
  author={Weitz-Shapiro, Rebecca and Winters, Matthew S},
  journal={Journal of Politics},
  volume={79},
  number={1},
  pages={60--74},
  year={2017},
  publisher={University of Chicago Press Chicago, IL}
}

@article{ferraz2008exposing,
  title={Exposing corrupt politicians: the effects of {B}razil's publicly released audits on electoral outcomes},
  author={Ferraz, Claudio and Finan, Frederico},
  journal={Quarterly Journal of Economics},
  volume={123},
  number={2},
  pages={703--745},
  year={2008},
  publisher={MIT Press}
}

@techreport{worldbank_2013,
	title = {Mapping context for social accountability : a resource paper},
	shorttitle = {Mapping context for social accountability},
	%url = {http://documents.worldbank.org/curated/en/293491468151492128/Mapping-context-for-social-accountability-a-resource-paper},
	abstract = {Demand-side governance and social accountability approaches (SAcc) have steadily gained prominence as a perceived means for achieving and improving a range of development outcomes. This resource paper focuses on the issue of SAcc and context, arising out of a growing recognition that context is critical in shaping, making, and breaking SAcc interventions. As such, the four main objectives of this paper are: to outline the main contextual factors that appear to be critical to SAcc; to examine how SAcc interventions interact with the context to bring about change in order to provide a preliminary, context-sensitive Theory of Change (ToC); to explore the operational implications that arise from first two objectives; and to offer a flexible analytical framework to guide practitioners wanting to undertake context analysis prior to engaging in demand-side activities. The paper attempted to achieve these objectives by: summarizing and building on a recently-conducted global review of the evidence-base, drawing on relevant conceptual literature to deepen understanding of SAcc and context, reviewing case-study material to extract indications of what types of SAcc approaches might work best when faced with different contextual realities, and holding consultations with experts and practitioners to test and modify the ideas being developed. The report is separated into five chapters and an annex. Chapter one introduces the topic and the rationale for undertaking work in this area. Chapter two outlines some of the key contextual variables that emerge as critical in shaping the form and effectiveness of SAcc. This provides a broad framework for understanding the important contextual constraints and opportunities. Chapter three outlines some of the key ways in which SAcc has influenced the context to produce positive change. Chapter four then explores and unpacks the practical implications of the approach. It offers two tools for SAcc practitioners to begin exploring ways to tailor to their contexts in a more structured manner. Finally, the annex, based on the paper's overall framework, provides a set of guiding questions for undertaking a context analysis prior to supporting SAcc operations.},
	language = {en},
	type={Working Paper},
	number = {79351},
	urldate = {2020-03-18},
	institution = {The World Bank},
	author = {O'Meally, Simon C.},
	month = jan,
	year = {2013},
	pages = {1--118}
}

@article{anwar2025efficient,
  title={Efficient public good provision between and within groups},
  author={Anwar, Chowdhury Mohammad Sakib and Bruno, Jorge and Foucart, Renaud and SenGupta, Sonali},
  journal={Games and Economic Behavior},
  volume={150},
  pages={183--190},
  year={2025},
  publisher={Elsevier}
}

@article{attanasi2019embezzlement,
  title={Embezzlement and guilt aversion},
  author={Attanasi, Giuseppe and Rimbaud, Claire and Villeval, Marie Claire},
  journal={Journal of Economic Behavior \& Organization},
  volume={167},
  pages={409--429},
  year={2019},
  publisher={Elsevier}
}

@article{battigalli2022belief,
  title={Belief-dependent motivations and psychological game theory},
  author={Battigalli, Pierpaolo and Dufwenberg, Martin},
  journal={Journal of Economic Literature},
  volume={60},
  number={3},
  pages={833--882},
  year={2022},
  publisher={American Economic Association 2014 Broadway, Suite 305, Nashville, TN 37203-2425}
}

@article{dufwenberg2004theory,
  title={A theory of sequential reciprocity},
  author={Dufwenberg, Martin and Kirchsteiger, Georg},
  journal={Games and economic behavior},
  volume={47},
  number={2},
  pages={268--298},
  year={2004},
  publisher={Elsevier}
}

@article{ellingsen2010testing,
  title={Testing guilt aversion},
  author={Ellingsen, Tore and Johannesson, Magnus and Tj{\o}tta, Sigve and Torsvik, Gaute},
  journal={Games and Economic Behavior},
  volume={68},
  number={1},
  pages={95--107},
  year={2010},
  publisher={Elsevier}
}

@article{besley2020state,
  title={State capacity, reciprocity, and the social contract},
  author={Besley, T.},
  journal={Econometrica},
  volume={88},
  number={4},
  pages={1307--1335},
  year={2020},
  publisher={Wiley Online Library}
}

@article{subrado2022moralagents,
  title={Taxing moral agents},
  author={Munoz Subrado, E.},
  journal={CESifo Working Paper No. 9867},
    year={2022}
}

@article{vanLeeuwenAlger2024,
  title={Estimating social preferences and Kantian morality in strategic interactions},
  author={van Leeuwen, B. and Alger, I.},
  journal={Journal of Political Economy: Microeconomics},
  volume={2},
  number={4},
  pages={665--706},
  year={2024},
  publisher={The univeristy of Chicago Press}
}

@article{EichnerPethig2021,
  title={Climate policy and moral consumers},
  author={Eichner, T. and Pethig, R.},
  journal={The Scandinavian Journal of Economics},
  volume={123},
  number={4},
  pages={1190--1226},
  year={2021},
  publisher={Wiley Online Library}
}

@article{AlmTorgler2011,
  title={So ethics matter? Tax compliance and morality},
  author={Alm, J. and Torgler, B.},
  journal={Journal of Business Ethics},
  volume={101},
  pages={635--651},
  year={2011}
}

@article{TianZhao2024,
  title={Tolerance for corruption and descriptive social norm: an experimental study of embezzlement},
  author={Tian, S. and Zhao, L.},
  journal={Plos One},
  volume={19},
number={5: e0303558},
  year={2024}
}

@article{MakowskyWang2018,
  title={Embezzlement, whistleblowing, and organizational architecture: an experimental investigation},
  author={Makowsky, M. D. and Wang, S.},
  journal={Journal of Economic Behavior and Organization},
  volume={147},
pages={58--75},
  year={2018}
}

@online{Reuters2025Mauritius,
  author    = {{Reuters}},
  title     = {Mauritius arrests ex-central bank governor, finance minister in embezzlement case},
  year      = {2025},
  month     = {4},
  day       = {9},
  url       = {https://www.reuters.com/world/africa/mauritius-arrests-ex-central-bank-governor-finance-minister-embezzlement-case-2025-04-09/},
  urldate   = {2025-08-22}
}

@article{CharnessDufenberg2006,
  title={Promises and partnership},
  author={Charness, G. and Dufwenberg, M.},
  journal={Econometrica},
  volume={74},
number={6},
pages={1579--1601},
  year={2006}
}

@article{BattigalliDufenberg2007,
  title={Guilt in games},
  author={Battigalli, P. and Dufwenberg, M.},
  journal={American Economic Review},
  volume={97},
number={2},
pages={170--176},
  year={2007}
}

@article{BenabouTirole2006,
  title={Incentives and Prosocial Behavior},
  author={Benabou, R. and Tirole, J.},
  journal={American Economic Review},
  volume={96},
number={5},
pages={1652-1678},
  year={2006}
}

@article{EllingsenJohannesson2008,
  title={Pride and Prejudice: The Human Side of Incentive Theory},
  author={Ellingsen, T. and Johannesson, M.},
  journal={American Economic Review},
  volume={98},
number={3},
pages={990--1008},
  year={2008}
}

@article{AnwarGeorgalos2024,
  title={Position uncertainty in a sequential public goods game: an experiment},
  author={Anwar, Chowdhury Mohammad Sakib and Georgalos, Konstantinos},
  journal={Experimental Economics},
  volume={27},
pages={820--853},
  year={2024}
}

@article{BattigalliDufwenberg2009,
  author  = {Battigalli, Pierpaolo and Dufwenberg, Martin},
  title   = {Dynamic Psychological Games},
  journal = {Journal of Economic Theory},
  year    = {2009},
  volume  = {144},
  number  = {1},
  pages   = {1--35}
}

\newpage

\appendix
\section*{Appendix A : Illustrative Example Proofs} \label{sec:AppA}

\iffalse

\subsubsection*{Proofs for Observations \ref{prop:FR_N2}--\ref{prop:mixed_N2}}

Throughout, fix $N=2$ and $k=1$. Each citizen chooses $c_i\in\{0,1\}$ in Stage 1. In Stage 2, exactly one citizen is audited, so $\Pr(\omega_1 = 1) = \Pr(\omega_2 = 1) = 1/2$. If $\omega_i=1$ and $c_i=0$, then citizen $i$ pays the contribution and penalty, i.e.\ pays $1+z$. If $c_i=1$, he pays $1$ regardless of $\omega_i$.

For any profile, the realized fund is $X=\max\{c_1,\omega_1\}+\max\{c_2,\omega_2\}\in\{1,2\}$.
Given contribution probabilities $(p_1,p_2)$, the Governor’s belief about the expected fund is
\[
\tau(p_1,p_2)=\mathbb{E}[X]=1+\frac{p_1+p_2}{2}.
\]
In deviations, the Governor’s belief $\tau(p_1,p_2)$ is determined by the prescribed equilibrium strategy profile and is therefore held fixed.

The citizen $i$'s payoff is
\[
u_i(c_i,\omega_i,g)= -\max\{c_i,\omega_i\}-(1-c_i)\omega_i z + ag,
\]
and the Governor’s payoff is
\[
u_G(g;X,\tau,\sigma)=(X-g)-\sigma \cdot \mathbf{1}\{g<\tau\}.
\]

\fi

\begin{proof}[Proof of Observation \ref{prop:FR_N2}]
Consider the symmetric profile with $p_1 = p_2 = 0$, i.e.\ $c_1=c_2=0$.

\paragraph{Governor’s best response.}
If $p_1 = p_2 = 0$, then $\tau(0, 0) = 1 = \mathbb{E}[X]$.
Given any realized common fund $X$, the Governor compares two relevant actions:
\[
g=0 \quad\text{yields}\quad u_G(0)=X-\sigma, 
\qquad
g=1 \ (\text{feasible if }X\ge1)\quad\text{yields}\quad u_G(1)=X-1.
\]
Thus, at $X=1$, she chooses $g=1$ if and only if $1-\sigma \le 0$, i.e.\ if $\sigma\ge1$, and otherwise chooses $g=0$.
Hence the stated Governor behavior is optimal.

\paragraph{Citizens’ deviation incentives.}
Fix the Governor behavior above and consider a unilateral deviation by citizen $i$ from $c_i=0$ to $c_i=1$.
Under the candidate equilibrium, the level of the common fund that the Governor believes citizens expect to observe remains $\tau(0, 0) = 1$.

\emph{Case 1: $\sigma\ge1$.} The Governor provides $g=1$ whenever feasible, and in particular provides $g=1$ both at $X=1$ (on the equilibrium path) and at $X=2$ (after a deviation).
Hence the deviation affects only citizen $i$’s private cost.
If citizen $i$ shirks ($c_i=0$), then with probability $1/2$ he is audited and pays $1+z$, and with probability $1/2$ he pays $0$, while he always enjoys benefit $ag$ with $g=1$:
\[
\mathbb{E}[u_i(0,\omega_i,1)]=\tfrac12(-1-z+a)+\tfrac12(0+a)= -\tfrac{1+z}{2}+a.
\]
If instead he contributes ($c_i=1$), he pays $1$ for sure and still obtains $g=1$:
\[
\mathbb{E}[u_i(1,\omega_i,1)]=-1+a.
\]
Thus shirking is optimal if and only if
\[
-\tfrac{1+z}{2}+a \ge -1+a \quad \Longleftrightarrow\quad z\le 1.
\]

\emph{Case 2: $\sigma<1$.} The Governor provides $g=0$ at both $X=1$ and $X=2$.
Again only private costs matter. Shirking yields expected payoff $-\tfrac12(1+z)$, and contributing yields $-1$, so no deviation occurs if
\[
-\tfrac12(1+z)\ge -1 \quad \Longleftrightarrow\quad z\le 1.
\]

Combining the Governor and citizen conditions, the profile $p_1 = p_2 = 0$ and $g=1$ is a SSPE if $z\le1$ and $\sigma \ge 1$, and 
the profile $p_1 = p_2 = 0$ and $g=0$ is a SSPE if $z\le1$ and 
$\sigma \le 1$.
\end{proof}

\begin{proof}[Proof of Observation \ref{prop:FC_N2}]
Consider the symmetric profile $p_1 = p_2 = 1$, i.e.\ $c_1=c_2=1$. Then 
$\tau(1, 1) = 2 = \mathbb{E}[X]$.

\paragraph{Governor’s best response.}
Given $\tau(1, 1) = 2 = \mathbb{E}[X]$, the Governor compares $g=2$ and $g=0$:
\[
u_G(2)= (2-2)-0=0,
\qquad
u_G(0)= (2-0)-\sigma=2-\sigma.
\]
Thus she chooses $g=2$ if $0\ge 2-\sigma$, or if $\sigma\ge2$; otherwise she chooses $g=0$.
Moreover, if the realized common fund $X=1$, then $g=2$ is infeasible, so the best feasible action is $g=0$.

\paragraph{Citizens’ deviation incentives.}
Fix the Governor behavior above and consider a unilateral deviation by citizen $i$ to $c_i=0$, holding the Governor belief $\mathbb{E}[X] = 2$ fixed.

\emph{Case (i): $\sigma\ge2$ (reputation-sensitive).}
On the equilibrium path, $g=2$ and citizen $i$’s payoff from contributing is
\[
\mathbb{E}[u_i(1,\omega_i,2)]=-1+2a.
\]
If citizen $i$ deviates to shirking, then:\\
- with probability $1/2$ he is audited, is forced to pay $1+z$, the fund remains $X=2$, and the Governor provides $g=2$;\\
- with probability $1/2$ he is not audited, the fund is $X=1<\mathbb{E}[X]$, so the Governor provides $g=0$.
Hence,
\[
\mathbb{E}[u_i(0,\omega_i,g')]=\tfrac12(-1-z+2a)+\tfrac12(0).
\]
No deviation requires
\[
-1+2a \ge \tfrac12(-1-z+2a)
\quad \Longleftrightarrow\quad
z \ge 1-2a.
\]

\emph{Case (ii): $\sigma<2$ (reputation-insensitive).}
The Governor provides $g=0$ regardless of $X\in\{1,2\}$.
Then, contributing yields $-1$, while shirking yields $-\tfrac12(1+z)$, so no deviation occurs if
\[
-1 \ge -\tfrac12(1+z)\quad \Longleftrightarrow\quad z\ge 1.
\]

This proves the two cases in the Observation 2. In \textit{case (i)} the outcome is efficient with $g=2$; in case (ii) full contribution is sustained by a large penalty but $g=0$.
\end{proof}

\begin{proof}[Proof of Observation \ref{prop:asym_N2}]
Consider the asymmetric profile $(c_1,c_2)=(1,0)$ and the associated level of the common fund that the Governor believes citizens expect to observe is

\[
\tau(1,0)=1+\frac{1+0}{2}=1.5.
\]
Given $\tau(1, 0) = 1.5$, the Governor’s best response is to provide $g=1.5$ if $X\ge1.5$ and $g=0$ if $X<1.5$, provided she is reputation-sensitive or  $\sigma\ge 1.5$.

Note that the realized common fund is $X=2$ if citizen 2 is audited (prob.\ $1/2$) and $X=1$ if citizen 1 is audited (prob.\ $1/2$). Therefore $g=1.5$ occurs when $X=2$ and $g=0$ occurs when $X=1$.

\paragraph{Payoffs.}
Citizen 1 pays $1$ for sure. Citizen 2 pays $0$ if not audited and $1+z$ if audited.
Thus,
\[
\mathbb{E}[u_1]=\tfrac12(-1+0)+\tfrac12(-1+1.5a)= -1+0.75a,
\]
\[
\mathbb{E}[u_2]=\tfrac12(0+0)+\tfrac12(-1-z+1.5a)= -\tfrac12(1+z)+0.75a.
\]

\paragraph{Deviations.}
We check unilateral deviations holding the Governor's belief $\tau(1, 0) = 1.5$ fixed.

If citizen 1 deviates to $c_1'=0$, then $X=1$ regardless of the audit (exactly one citizen is audited and citizen 2 shirks), so $X<1.5$ and hence $g=0$. His deviation payoff is
\[
\mathbb{E}[u_1(0,\omega_1,0)]=\tfrac12(-1-z)+\tfrac12(0)= -\tfrac12(1+z).
\]
No deviation for citizen 1 requires
\[
-1+0.75a \ge -\tfrac12(1+z)\quad \Longleftrightarrow\quad z \ge 1-1.5a.
\]

If citizen 2 deviates to $c_2'=1$, then $X=2$ regardless of the audit, so $X\ge1.5$ and $g=1.5$. His deviation payoff is
\[
\mathbb{E}[u_2(1,\omega_2,1.5)]=-1+1.5a.
\]
No deviation for citizen 2 requires
\[
-\tfrac12(1+z)+0.75a \ge -1+1.5a\quad \Longleftrightarrow\quad z \le 1-1.5a.
\]
Thus both no-deviation inequalities hold if and only if $z = 1-1.5a$, and the Governor condition requires $\sigma\ge1.5$.
This establishes the knife-edge equilibrium claim.
\end{proof}

\begin{proof}[Proof of Observation \ref{prop:mixed_N2}]
We derive mixed-strategy SSPE for $N=2,k=1$.

\paragraph{Case (i): $\sigma\ge 2$.}
If $\sigma\ge 2$, then for any $p\in(0,1)$ we have $\tau(p)=1+p\le 2\le \sigma$, so the Governor is reputation-sensitive for all relevant $p$.
Hence, given $p$, she provides
\[
g=
\begin{cases}
\tau(p)=1+p, & \text{if } X\ge 1+p,\\
0, & \text{if } X<1+p.
\end{cases}
\]
When a citizen $i$ is not audited (prob.\ $1/2$), if he contributes then $X=2$ and thus $g=1+p$; if he shirks then $X=1$ and thus $g=0$.
When citizen $i$ is audited (prob.\ $1/2$), the fund is $X=2$ with probability $p$, or $X=1$ with probability $1-p$.

Therefore, the expected payoff from contributing is
\[
\mathbb{E}[u_i(1,\omega_i,\cdot)] = -1 + \frac{1}{2}a(1+p)+ \frac{1}{2}ap(1+p),
\]
while the expected payoff from shirking is 
\[
\mathbb{E}[u_i(0,\omega_i,\cdot)] = \frac12(-1-z+ap(1+p)) + \tfrac12(0).
\] 
Indifference requires 
\[
-1 + \frac{1}{2}a(1+p)+ \frac{1}{2}ap(1+p) \;=\; \frac12(-1-z+ap(1+p)),
\]
which simplifies to
\[
a(1+p) = 1-z
\quad \Longleftrightarrow\quad
p^*=\frac{1-z}{a}-1.
\]
A mixed strategy requires $0<p^*<1$, i.e.
\[
0 < \frac{1-z}{a}-1 < 1
\quad \Longleftrightarrow\quad
1-2a < z < 1-a,
\]
with $a>0$ (and in the model $a\in(0,1)$).
Uniqueness follows because the indifference equation is linear in $p$ in this case.

\paragraph{Case (ii): $\sigma<1$.}
If $\sigma<1$, then for any $p\in(0,1)$ we have $\tau(p)=1+p>1>\sigma$, so the Governor is always reputation-insensitive and sets $g=0$ for all $X$.
Hence contributing yields payoff $-1$, while shirking yields expected payoff $-\tfrac12(1+z)$.
Indifference requires
\[
-1 = -\tfrac12(1+z)\quad \Longleftrightarrow\quad z=1.
\]
If $z=1$ and $\sigma<1$, then every $p\in(0,1)$ satisfies the indifference condition and therefore any such $p$ constitutes a mixed-strategy SSPE with $g=0$.

\end{proof}

\section*{Appendix B : Example: $N=3, k=1$}
\label{sec:AppB}
%In this section, we analyze the non-trivial setting with three citizens, $N = 3$, and one audit, $k=1$. 

%\subsection{Example: $N=3, k=1$ }

We now analyse the case with three citizens, $N=3$, and one audit, $k=1$. 

\paragraph{Citizens' Expectations}
If each citizen contributes with probability $p$, the expected size of the common fund is
\begin{equation}
\tau(p) = k + p(N-k) = 1 + 2p.
\end{equation}

\paragraph{Reputation-Sensitive Governor}
Suppose the Governor's reputation sensitivity satisfies $\sigma \ge \tau(p)$ for the relevant range of $p$. Since $\tau(p) \le 3$ on $(0,1)$, this holds if $\sigma \ge 3$. In this regime, she aims to meet expectations whenever feasible, setting $g=\tau(p)$ if the realised fund $X$ reaches the expectation threshold for that $p$, and $g=0$ otherwise.

We analyse two cases.

\medskip
\noindent\textbf{Case A: $0 < p \le \tfrac{1}{2}$.} \\
Here $\tau(p)\le 2$, so the Governor provides $g=\tau(p)$ whenever $X \ge 2$, and $g=0$ otherwise.

Consider a citizen $i$. If he is \emph{not} audited (probability $2/3$), then:

\begin{itemize}
\item If citizen $i$ contributes, there is always at least the audited citizen (among the other two) plus $i$ himself, so $X \ge 2$ and $g=\tau(p)$ for sure.
\item If citizen $i$ shirks, there is still one audited contributor among the other two citizens, but to reach $X\ge 2$ the remaining non-audited other citizen must also contribute voluntarily (probability $p$). Thus $g=\tau(p)$ with probability $p$ and $g=0$ with probability $1-p$.
\end{itemize}

When citizen $i$ is audited (probability $1/3$), neither of the other two citizens is audited, so both contribute independently with probability $p$. Therefore, the fund is $X = 1$ plus the number of voluntary contributors among the other two citizens. Hence $X \geq 2$ and $g = \tau(p)$ with probability $1-(1-p)^2 = p(2-p)$, and $g = 0$ otherwise.

The expected utilities are therefore 
\begin{align*}
    \mathbb{E}[u_i(1,\omega_i,\cdot)]
    &= \tfrac{2}{3}\bigl[-1 + a(1+2p)\bigr] + \tfrac{1}{3}\bigl[-1 + p(2-p)\,a(1+2p)\bigr], \\[4pt]
    \mathbb{E}[u_i(0,\omega_i,\cdot)]
    &= \tfrac{2}{3}\bigl[p\,a(1+2p)\bigr] + \tfrac{1}{3}\bigl[-(1+z) + p(2-p)\,a(1+2p)\bigr].
\end{align*}
Setting these expressions equal, the $p(2-p)\,a(1+2p)$ terms cancel out and the indifference condition simplifies to the quadratic equation

\begin{equation}
2a p^2 - a p + \left(1 - a - \frac{z}{2}\right) = 0,
\qquad 0 < p \le \frac{1}{2}.
\label{eq:caseAquad}
\end{equation}
A mixed-strategy SSPE exists in Case A if and only if \eqref{eq:caseAquad} has a real root in $(0,\tfrac{1}{2}]$.

\medskip
\noindent\textbf{Case B: $\tfrac{1}{2} < p < 1$.} \\
Here $\tau(p) > 2$, so the Governor provides $g=\tau(p)$ only if $X \ge 3$ (i.e., all three citizens contribute).

Consider the not-audited citizen $i$ (probability $2/3$):

\begin{itemize}
\item If citizen $i$ contributes, we reach $X=3$ only if both other citizens contribute, in which case $g=\tau(p)$; otherwise $g=0$.
\item If citizen $i$ shirks, $X=3$ is impossible (at least one citizen, $i$, does not contribute), so $g=0$.
\end{itemize}

When citizen $i$ is audited (probability $1/3$), neither of the other two citizens is audited. The fund is $X = 3$ and $g = \tau(p)$ only if both other citizens contribute voluntarily (probability $p^2$), and $g = 0$ otherwise.

The expected utilities are therefore
\begin{align*}
    \mathbb{E}[u_i(1,\omega_i,\cdot)]
    &= \tfrac{2}{3}\bigl[-1 + p\,a(1+2p)\bigr] + \tfrac{1}{3}\bigl[-1 + p^2\,a(1+2p)\bigr], \\[4pt]
    \mathbb{E}[u_i(0,\omega_i,\cdot)]
    &= \tfrac{2}{3}(0) + \tfrac{1}{3}\bigl[-(1+z) + p^2\,a(1+2p)\bigr].
\end{align*}
Setting these expressions equal, the $p^2\,a(1+2p)$ terms cancel out and the indifference condition becomes

\begin{equation}
2a p^2 + a p - \left(1 - \frac{z}{2}\right) = 0,
\qquad \frac{1}{2} < p < 1.
\label{eq:caseBquad}
\end{equation}
A mixed-strategy SSPE exists in Case B if and only if \eqref{eq:caseBquad} has a real root in $(\tfrac{1}{2},1)$.

\paragraph{Multiplicity under high reputation sensitivity}
When $\sigma \ge 3$, one or both of the equations \eqref{eq:caseAquad} and \eqref{eq:caseBquad} may admit roots in their respective intervals, so depending on $(a,z)$ there may be one or multiple mixed-strategy SSPE.

\paragraph{Reputation-Insensitive Governor}
If $\sigma < 1$, then $\sigma < \tau(p)$ for all $p \in (0,1)$ and the Governor always provides $g=0$. A citizen who contributes pays $1$ for sure; a shirker is audited with probability $1/3$ and then pays $1+z$. Indifference requires
\begin{equation}
-1 = -\frac{1}{3}(1+z) \quad \Longrightarrow \quad z = 2.
\end{equation}
Hence, if $\sigma < 1$ and $z=2$, any $p \in (0,1)$ is a symmetric mixed-strategy SPE with $g=0$. For $z \neq 2$, no interior mixed equilibrium exists in this low-$\sigma$ regime.

\paragraph{Summary}
\begin{itemize}
\item \textbf{High reputation sensitivity} ($\sigma \ge 3$): 
\begin{itemize}
\item Case A ($0 < p \le \tfrac{1}{2}$): Mixed-strategy SSPE exists if $2a p^2 - a p + (1 - a - z/2) = 0$ admits a root in $(0,\tfrac{1}{2}]$.

    \[
    2 - \frac{9}{4}a \le z \le 2-2a \quad \text{and} \quad 0<a\leq 1.
    \]

\item Case B ($\tfrac{1}{2} < p < 1$): Mixed-strategy SSPE exists if $2a p^2 + a p - (1 - z/2) = 0$ admits a root in $(\tfrac{1}{2},1)$.

  \[
    2 - 6a < z < 2-2a  \quad \text{and} \quad 0<a \leq 1.
    \]
\end{itemize}
Both cases can hold simultaneously for some $(a,z)$, yielding multiple mixed-strategy SSPE.
\item \textbf{Low reputation sensitivity} ($\sigma < 1$): a continuum of mixed-strategy SSPE (any $p \in (0,1)$ and $g=0$) exists if $z=2$.

\end{itemize}

\begin{figure}[htbp]
    \centering
    \begin{subfigure}[b]{0.45\textwidth}
        \centering
        \includegraphics[width=\textwidth]
       {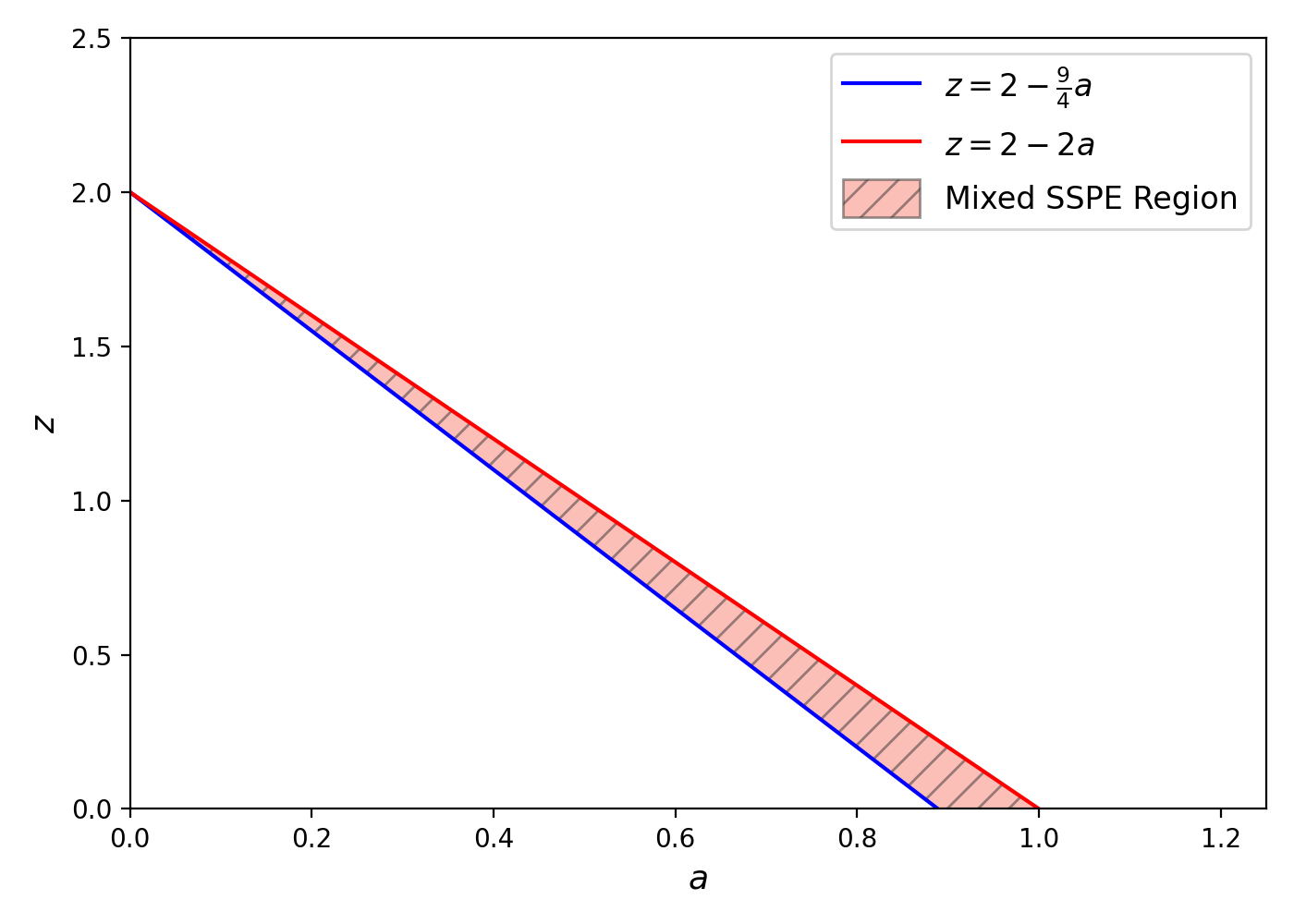}
            \caption{Feasible region for mixed-strategy SSPE with $0 < p \leq \tfrac{1}{2}$, for $N=3$, $k=1$, $\sigma \geq 3$. The shaded region is bounded by $z = 2 - \tfrac{9}{4}a$ (blue) and $z = 2 - 2a$ (red). A mixed-strategy SSPE in Case~A exists if and only if $(a,z)$ lies within this region.}

        \label{fig:N3_first}
    \end{subfigure}
    \hfill
    \begin{subfigure}[b]{0.45\textwidth}
        \centering
        \includegraphics[width=\textwidth] {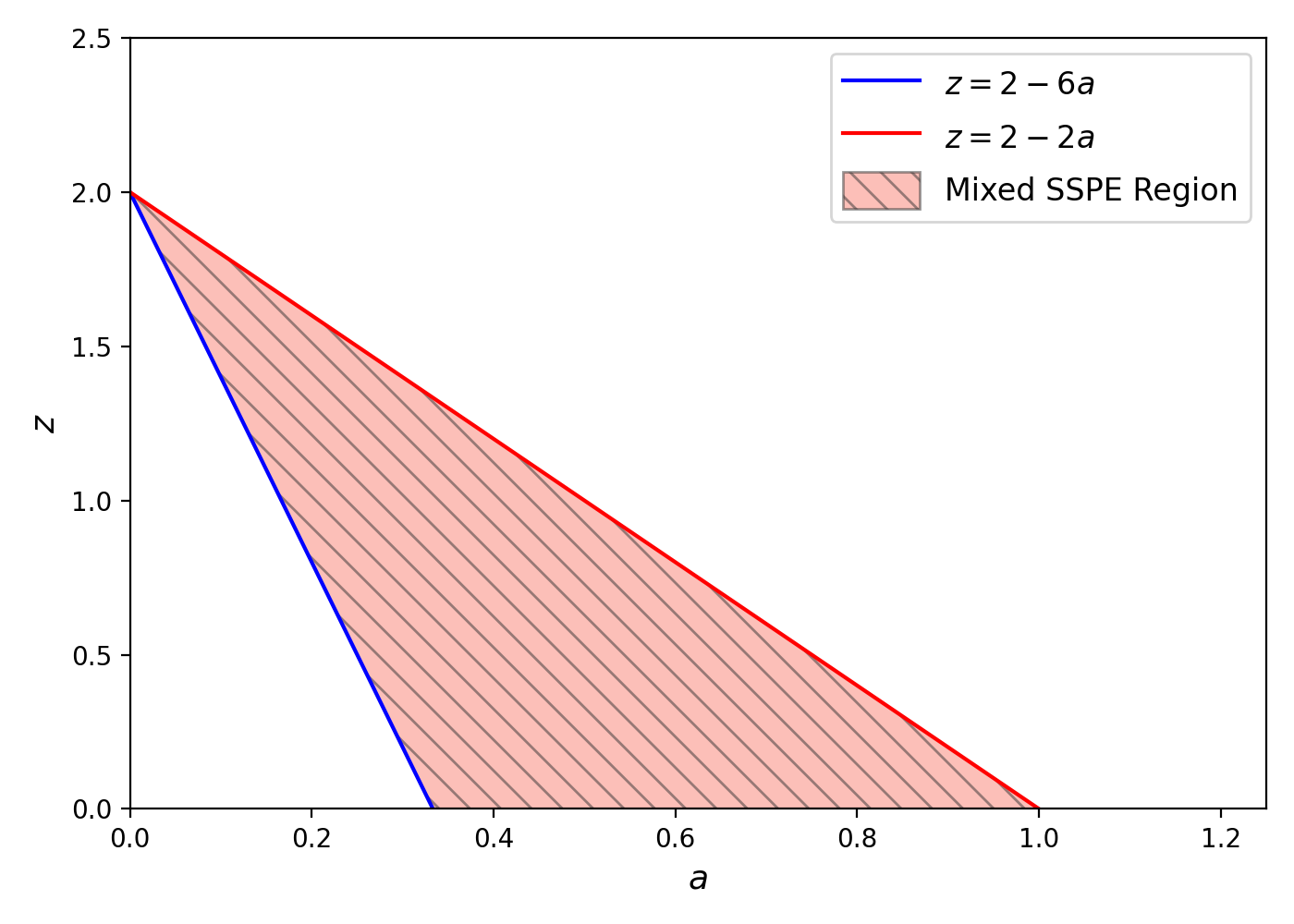}
            \caption{Feasible region for mixed-strategy SSPE with $\tfrac{1}{2} < p < 1$, for $N=3$, $k=1$, $\sigma \geq 3$. The shaded region is bounded by $z = 2 - 6a$ (blue, dashed) and $z = 2 - 2a$ (red). A mixed-strategy SSPE in Case~B exists if and only if $(a,z)$ lies within this region.}

        \label{fig:N3_second}
    \end{subfigure}
    \caption{Feasible Regions for $N=3,k=1$}
    \label{fig:N3_regions}
\end{figure}

\begin{figure}[h!]
    \centering
    \includegraphics[width=0.55\linewidth]{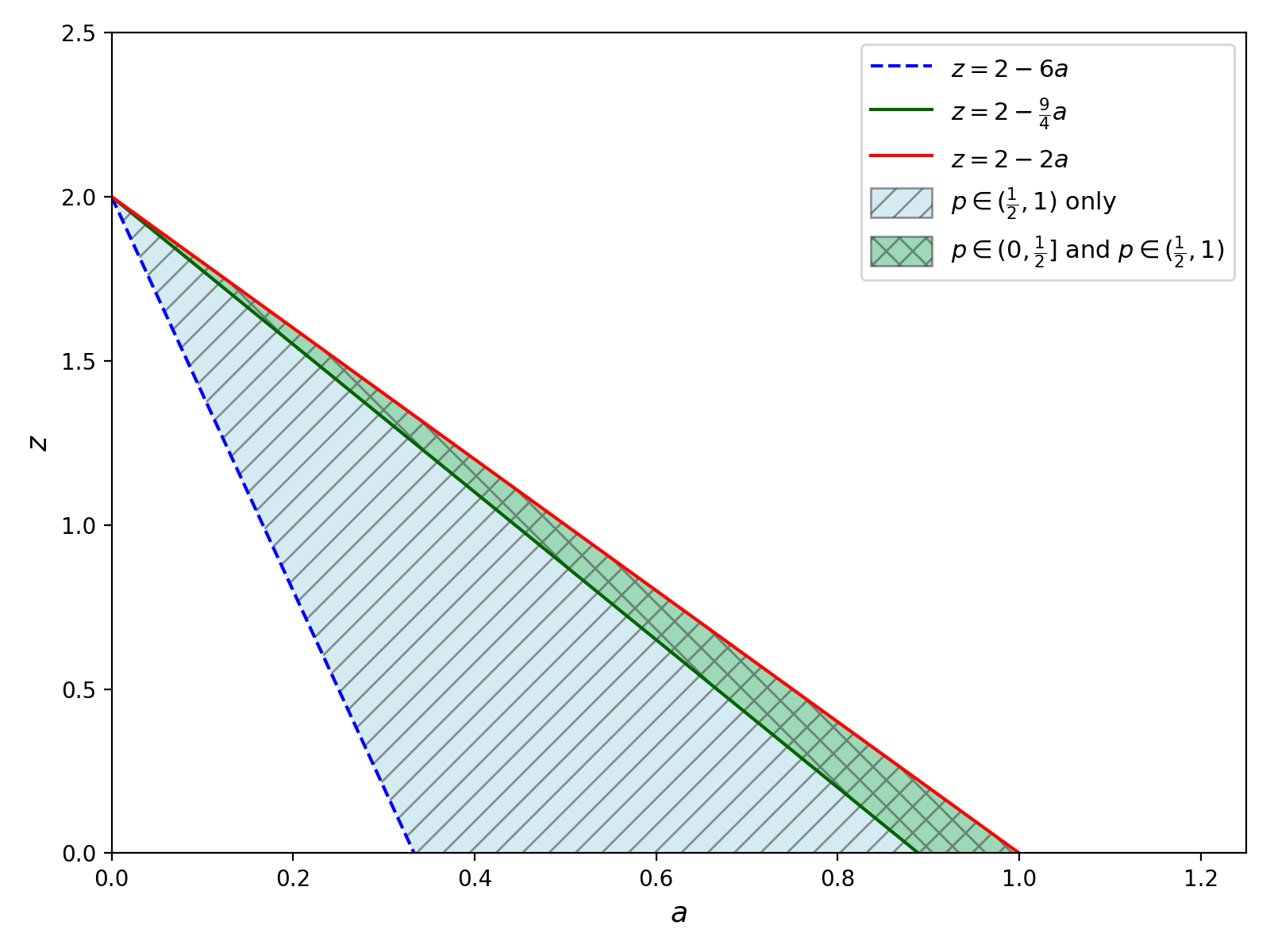}
       \caption{Combined feasible region for all mixed-strategy SSPE with $p \in (0,1)$, for $N=3$, $k=1$, $\sigma \geq 3$. The light blue region (between $z = 2 - 6a$ and $z = 2 - \tfrac{9}{4}a$) admits only Case~B equilibria with $p \in (\tfrac{1}{2},1)$. The green region (between $z = 2 - \tfrac{9}{4}a$ and $z = 2 - 2a$) admits both Case~A equilibria with $p \in (0,\tfrac{1}{2}]$ and Case~B equilibria with $p \in (\tfrac{1}{2},1)$, yielding multiple mixed-strategy SSPE. No mixed-strategy SSPE exists outside the shaded regions.}

    \label{fig:N3_combined}
\end{figure}

Figures \ref{fig:N3_regions} and \ref{fig:N3_combined} show that mixed-strategy SSPE exist within a narrow parameter region. In that region, there may be either a unique or multiple mixed-strategy SSPE.

\newpage

\phantomsection
\section*{Appendix C : Proofs}
\label{AppC}
\begin{proof}[Proof of Lemma \ref{lemma1}]

For each $i$, recall that $\omega_i\in\{0,1\}$ indicates whether $i$ is audited, and $c_i\in\{0,1\}$ indicates voluntary contribution. By the rule that audited citizens are compelled to pay, the realised contribution from $i$ is
\[
\max\{c_i,\omega_i\} \;=\; \omega_i + (1-\omega_i)c_i.
\]
Hence, the total common fund is
\[
X \;=\; \sum_{i=1}^N \max\{c_i,\omega_i\}
   \;=\; \sum_{i=1}^N \bigl[\omega_i + (1-\omega_i)c_i\bigr].
\]
Taking expectations and using linearity,
\[
\mathbb{E}[X]
= \sum_{i=1}^N \Big( \mathbb{E}[\omega_i] + \mathbb{E}[(1-\omega_i)c_i] \Big).
\]
Citizens choose $c_i$ before the audit and the audit is random; thus $c_i$ is independent of $\omega_i$: 
\[
\mathbb{E}[(1-\omega_i)c_i] \;=\; \mathbb{E}[1-\omega_i]\;\mathbb{E}[c_i]
\;=\; \Big(1-\frac{k}{N}\Big)\,p.
\]
Since exactly $k$ of the $N$ citizens are audited uniformly at random, for each $i$ we have
\[
\mathbb{E}[\omega_i] \;=\; \Pr(\omega_i=1) \;=\; \frac{k}{N}.
\]
Therefore, for every $i$,
\[
\mathbb{E}\big[\max\{c_i,\omega_i\}\big]
\;=\; \frac{k}{N} + \Big(1-\frac{k}{N}\Big)p,
\]
and summing over $i=1,\dots,N$ yields
\[
\mathbb{E}[X]
\;=\; N\left[\frac{k}{N} + \Big(1-\frac{k}{N}\Big)p\right]
\;=\; k + (N-k)p.
\]
\end{proof}

\begin{proof}[Proof of Lemma \ref{lemma:continuity}]

To prove that the function $f(p)$ is continuous on the entire interval $(0,1)$, we first show it is continuous on any open interval between the kink points, and then show that the one-sided limits are equal at every kink point.

\paragraph{Continuity between kink points.}
Let the set of critical points (the "kinks") be $\{p_m \mid \tau(p_m)=m \in \mathbb{Z}, \text{ for } k<m<N\}$. This is a finite set of points that partitions the interval $(0,1)$ into a finite number of open subintervals of the form $(p_m, p_{m+1})$. Consider any such open subinterval. For any $p$ within this interval, the integer threshold for the common fund, $m(p)=\lceil\tau(p)\rceil$, is a constant integer. As a result, the number of pivotal contributors needed, $j(p)=m(p)-1-k$, is also a constant number for this entire interval.  Let's call it $J$. Also, note that the term $\left(1 - \frac{k z}{N-k}\right)$ is a constant. For this interval, the general formula for $f(p)$:
\[
f(p)=a \cdot \tau(p) \cdot\binom{N-1-k}{j(p)} p^{j(p)}(1-p)^{N-1-k-j(p)}-\left(1 - \frac{k z}{N-k}\right),
\]
simplifies to:
\[
f(p)=a \cdot(k+p(N-k)) \cdot\left[\binom{N-1-k}{J} p^J(1-p)^{N-1-k-J}\right]-C.
\]
Since $J$ is a constant, the term in the square brackets is a polynomial in $p$. The entire expression is a linear function multiplied by a polynomial, which results in another polynomial. The difference between this polynomial and the constant $C$ is also a polynomial. A polynomial is continuous everywhere. Therefore, the function $f(p)$ is continuous on every open interval between the kink points.

\paragraph{Continuity at the kink points.} Now, we only need to show that the function is also continuous at the kink points themselves by proving that the limit from the left is equal to the limit from the right.

Let $p_m$ be an arbitrary kink point, defined by the condition that the Governor's threshold crosses an integer $m$:
\[
\tau(p_m) = k + p_m(N-k) = m.
\]

This implies that the location of the kink is $p_m = \frac{m-k}{N-k}$. The Net Gain function is:

\[
f(p) = a \cdot \tau(p) \cdot \binom{N-1-k}{j(p)} p^{j(p)} (1-p)^{N-1-k-j(p)} - C,
\]
where $j(p) = \lceil\tau(p)\rceil - 1 - k$ is the number of pivotal contributors and $C = \left(1 - \frac{k z}{N-k}\right)$ is the constant net marginal cost.

We first evaluate the limit as $p$ approaches $p_m$ from below ($p \to p_m^-$). As $p \to p_m^-$, the threshold $\tau(p)$ approaches $m$ from below, so the integer threshold is $\lceil\tau(p)\rceil = m$. The number of pivotal contributors is $j = m - 1 - k$. The left-hand limit is:
\[
\lim_{p \to p_m^-} f(p) = a \cdot \tau(p_m) \cdot \binom{N-1-k}{m-1-k} p_m^{m-1-k} (1-p_m)^{N-m} - C.
\]

Next, we evaluate the limit as $p$ approaches $p_m$ from above ($p \to p_m^+$). As $p \to p_m^+$, the threshold $\tau(p)$ approaches $m$ from above, so the integer threshold is $\lceil\tau(p)\rceil = m+1$. The number of pivotal contributors is now $j = (m+1) - 1 - k = m-k$. The right-hand limit is:
\[
\lim_{p \to p_m^+} f(p) = a \cdot \tau(p_m) \cdot \binom{N-1-k}{m-k} p_m^{m-k} (1-p_m)^{N-m-1} - C.
\]

The function is continuous at $p_m$ if and only if the left-hand limit equals the right-hand limit. Since the terms $a \cdot \tau(p_m)$ and $C$ are the same for both limits, we only need to prove that the two binomial probabilities are equal at the point $p=p_m$:
\[
\binom{N-1-k}{m-1-k} p_m^{m-1-k} (1-p_m)^{N-m} = \binom{N-1-k}{m-k} p_m^{m-k} (1-p_m)^{N-m-1}
\]
Let's analyze the ratio of the right-hand side to the left-hand side:
\[
\frac{\binom{N-1-k}{m-k}}{\binom{N-1-k}{m-1-k}} \cdot \frac{p_m^{m-k}}{p_m^{m-1-k}} \cdot \frac{(1-p_m)^{N-m-1}}{(1-p_m)^{N-m}}
\]
This simplifies to:
\[
\left(\frac{N-m}{m-k}\right) \cdot (p_m) \cdot \left(\frac{1}{1-p_m}\right) = \left(\frac{N-m}{m-k}\right) \cdot \left(\frac{p_m}{1-p_m}\right).
\]
Now, we substitute the value of $p_m = \frac{m-k}{N-k}$:
\[
\frac{p_m}{1-p_m} = \frac{\frac{m-k}{N-k}}{1-\frac{m-k}{N-k}} = \frac{\frac{m-k}{N-k}}{\frac{N-k-(m-k)}{N-k}} = \frac{m-k}{N-m}.
\]
Substituting this back into our ratio gives:
\[
\left(\frac{N-m}{m-k}\right) \cdot \left(\frac{m-k}{N-m}\right) = 1.
\]
Since the ratio of the two probabilities is exactly 1, the probabilities are equal. Therefore, the left-hand limit equals the right-hand limit.

We have shown that the function $f(p)$ is continuous on any open interval between the kink points, and we have shown that the function $f(p)$ is continuous on all open intervals between the kink points and that the one-sided limits are equal at every kink point. Therefore, the function $f(p)$ is continuous on the entire interval $(0,1)$. This completes the proof.
\end{proof}

\begin{proof}[Proof of Proposition \ref{Thoerem: 4 Existence>N}]

%By Lemma \ref{lemma:continuity}, $f(p)$ is continuous on the interval $(0,1)$. 
In the mixed-strategy SSPE, each citizen contributes with probability $p$, where $p$ is the solution of the following equation $f(p) = 0$.
To prove that such a root exists, we establish a set of sufficient conditions under which the function is guaranteed to cross the $x$-axis. We do this by demonstrating that for a specific range of parameters, $f(p)$ is negative as $p \rightarrow 0$ and positive as $p \rightarrow 1$. Since function $f(p)$ is continuous and starts negative and ends positive, the Intermediate Value Theorem guarantees that it must cross the $x$-axis at least once. %Therefore, a point $p^* \in(0,1)$ where an equilibrium exists is guaranteed.
%A mixed-strategy equilibrium is a root of the Net Gain function, $f(p)$. We have also proven in Lemma \ref{lemma:continuity} that $f(p)$ is continuous on the interval $(0,1)$. 

\paragraph{Analysis of the Lower Boundary ($p \to 0$).}
We evaluate the limit of $f(p)$ as $p$ approaches 0 from the right. For the function to start negative, we require this limit to be less than zero. As shown in the analysis of the free-riding SSPE, the expected public good gain approaches $ak$. The condition is therefore:
\[
\lim_{p\to 0^+} f(p) < 0 \implies ak - \left(1 - \frac{kz}{N-k}\right) < 0 \implies z < \frac{N-k}{k}(1-ak).
\]

\paragraph{Analysis of the Upper Boundary ($p \to 1$).}
We evaluate the limit of $f(p)$ as $p$ approaches 1 from the left. For the function to end positive, we require this limit to be greater than zero. As shown in the analysis of the full-contribution SSPE, the expected public good gain approaches $aN$. The condition is therefore:
\[
\lim_{p\to 1^-} f(p) > 0 \implies aN - \left(1 - \frac{kz}{N-k}\right) > 0 \implies z > \frac{N-k}{k}(1-aN).
\]

The boundary analysis provides a set of sufficient conditions under which a mixed-strategy SSPE is guaranteed to exist. These conditions define a non-empty ``feasibility region'' in the $(a,z)$ parameter space. The conditions are as follows:
\begin{enumerate}[label=(\roman*)]
    \item \textbf{Condition on $a$:} The citizen's marginal per capita return of the public good must be positive but not excessively large. This condition arises because the penalty $z$ must be non-negative, which requires the upper bound for $z$ to be positive. Formally:
    \[
    \frac{N-k}{k}(1-ak) > 0.
    \]
    Since $N>k$, the term $\frac{N-k}{k}$ is positive. The inequality thus simplifies to $1-ak > 0$, which implies $a < 1/k$. Combined with the model's initial assumption that $a>0$, the condition is:
    \[
    0 < a < \frac{1}{k}.
    \]
    \item \textbf{Condition on $z$:} Given a value of $a$ that satisfies the above condition, the penalty $z$ must be non-negative and fall within the window defined by the boundary analysis.
    \[
    \max\left\{0, \frac{N-k}{k}(1-aN)\right\} \le z < \frac{N-k}{k}(1-ak).
    \]
\end{enumerate}

For any pair of parameters $(a, z)$ satisfying these two conditions, we have established that $\lim _{p \rightarrow 0^{+}} f(p)<0$ and $\lim _{p \rightarrow 1^{-}} f(p)>0$. Since function $f(p)$ is continuous, the Intermediate Value Theorem guarantees that there must exist at least one root $p^* \in(0,1)$ such that $f\left(p^*\right)=$ 0 . The fact that the upper bound for this window on $z$ is always greater than the lower bound (since $a N>a k$ ) ensures that this feasibility region is always non-empty for any $a \in(0,1 / k)$. This completes the proof.
\end{proof}

\end{document}